\documentclass[a4paper, 12pt]{article}
\usepackage[utf8]{inputenc}
\usepackage[english]{babel}
\usepackage[T1]{fontenc}
\usepackage{amsmath}
\usepackage{amssymb}
\usepackage{graphicx}
\usepackage{lscape}
\usepackage{listings}
\usepackage{float}
\usepackage{physics}
\usepackage[style=authoryear, maxcitenames = 2, maxbibnames=10, backend=bibtex]{biblatex}
\renewbibmacro{in:}{\ifentrytype{article}{}{\printtext{\bibstring{in}\intitlepunct}}}
\AtBeginDocument{}
\usepackage{url}
\usepackage{multicol}
\usepackage{xcolor}
\usepackage{bbm}

\usepackage{setspace}
\usepackage[left=2cm,right=2cm,top=3cm,bottom=3cm]{geometry}
\usepackage{indentfirst}
\usepackage{color, soul}

\usepackage{fancyhdr}
\usepackage{draftwatermark}
\SetWatermarkText{Working paper}
\SetWatermarkScale{0.5}
\SetWatermarkAngle{45}

\begin{document}

\newpage
\pagenumbering{arabic}
\lhead{Price of liquidity in the reinsurance of fund returns}
\chead{}
\rhead{\slshape --\thepage --}
\cfoot{}
\renewcommand{\headrulewidth}{0.4pt}
\begin{titlepage}

\begin{center}
\vspace*{1cm}
{\Huge {\bf Price of liquidity in the reinsurance of fund returns}}\\
\vspace*{2.5cm}
{David Saunders} \\ 
{\it Department of Statistics and Actuarial Science, University of Waterloo, Waterloo, Canada} \\
{\it dsaunders@uwaterloo.ca} \linebreak\\
{Luis Seco} \\ 
{\it CEO Sigma Analysis \& Management and Department of Mathematics, University of Toronto, Toronto, Canada} \\
{\it seco@math.utoronto.ca} \linebreak\\
{Markus Senn} \\ 
{\it Department of Mathematical Finance, Technical University Munich, Munich, Germany} \\
{\it markus.senn@tum.de} \\
\end{center}

\textbf{Abstract}

\textit{This paper aims to extend downside protection to a hedge fund investment portfolio based on shared loss fee structures that have become
increasing popular in the market. In particular, we consider a second tranche and suggest the purchase of an upfront reinsurance contract for any losses on the fund beyond the threshold covered by the first tranche, i.e. gaining full portfolio protection. We identify a fund's underlying liquidity as a key parameter and study the pricing of this additional reinsurance using two approaches: First, an analytic closed-form solution based on the Black-Scholes framework and second, a numerical simulation using a Markov-switching model. In addition, a simplified backtesting method is implemented to evaluate the practical application of the concept.} 

\vspace*{0.5cm}

\end{titlepage}

\newpage
\section{Introduction}
The low interest rate environment, uncertain and fragile political climate and cost pressure are main drivers for today's challenging market environment. It is increasingly difficult for hedge fund managers to justify high fees for return performances that tend not to be superior to alternative low-cost investment opportunities, e.g. ETFs. A hedge fund's capability of deliberately incorporating more risk by either adding new risk sources or including riskier positions poses an attractive alternative for investors to diversify their existing investment portfolio. Nonetheless, weak performance, more difficult and volatile markets, and pressure to raise assets combined with the investor's call for better alignment of interest have resulted in new fee structures in the hedge fund industry. Managers offer downside protection by insuring investments in the hedge fund in return for a bigger share in generated profits. These fee structures are commonly referred to as `first-loss', `shared-loss' or `shared-profit'.

In traditional fee structures the manager receives a flat 2\% management fee and a performance fee of 20\% of net profits (known as `2 and 20'). Even though the `2 and 20' concept is commonly applied, other fixed and variable fee combinations are possible, e.g. \textcite{hoehn} suggest an optimal equilibrium in management and performance fees by optimizing the commonly used Sharpe-Ratio to maximize the utility of both the manager and investor alike. Nonetheless new fee structures in the form of first-loss schemes are emerging. There are several versions of the first-loss principle, but the basic framework has the following features in common: the manager provides downside protection by covering the first tranche of portfolio losses in return for a higher participation rate in case of a positive fund performance. In contrast to traditional fee structures, a first-loss structure might allow a manager to raise the performance fee to 50\% by insuring losses up to 10\% of the initial investment, i.e. providing protection for the first tranche of portfolio losses. \textcite{hekou} derive trading strategies with a first-loss structure based on cumulative prospect theory and conclude that certain parameter combinations result in increasing utility for both the manager and the investor while reducing the risk of an investment portfolio. 
However, in other cases, the traditional fee structure produces higher investor utility. 
\textcite{secosharedloss} price the contract using risk-neutral valuation, thereby obtaining the `fair' level of the performance fee.
\textcite{fixedincomereturns} extend the traditional first-loss principle by introducing a minimum return guarantee in addition to covering the first tranche of losses in a portfolio. The investment resembles a coupon bond, with a guaranteed minimum return (fixed part) and performance dependent return net of accrued fees (variable part). 

Independent of applied fee structures, hedge funds are often subject to so-called liquidation barriers. These barriers can be introduced exogenously, i.e. they are investor demanded or endogenously, i.e. the manager sets an internal barrier. They represent a commonly known option to protect investors from negative fund performances. Once poorly performing funds breach the liquidation barrier, their operations shut down to prevent further losses for the investors. \textcite{jackwerth} analyze the effects of a shutdown or liquidation barrier on the risk-taking behavior of the manager as another option to provide downside protection for the investor. They find a significant change in the willingness to take on riskier positions when the fund performance is close to the barrier. 

Our approach aims to extend the protection beyond the first tranche of losses by considering a second one that insures the investor against all losses, not just the first level. The interest in this second layer of protection is manifold. On the one hand, investors may have a different appetite for the second layer of losses from a third-party reinsurer, therefore creating a financial transaction between investor and reinsurer. On the other hand, even if investors' risk appetites lead them to accept this second tranche of losses, or if reinsurance cost is deemed too high, there may be regulatory reasons for this tranche to be outsourced to a third party, e.g., if the investor is an insurance company subject to capital ratios linked to a loss profile that regulators may penalize in excess of the external reinsurance cost. In any case, whilst most related research focuses on optimizing fee structures in a first-loss framework, little work has been done on protecting losses greater than the first tranche. We address the subject of providing additional downside protection that goes beyond the insurance of the first tranche of losses.\footnote{A similar payoff in a 
different financial contract is afforded to investors in variable annuities, who may receive a minimum guaranteed rate of return, together with fractional participation in the 
upside of an investment portfolio managed by an insurance company.} 

Theoretically, by introducing a liquidation barrier at the lower end of the first tranche, the investor gains full portfolio protection. In this setting, the liquidation is triggered as soon as the portfolio value breaches the barrier, in which case the manager compensates for all resulting losses. In practice however, the investor is exposed to gap risk, arising from the inability of the manager of the investment structure to liquidate the portfolio before losses exceed the protection level of the first tranche. The risk premium that is relevant in the valuation of the second loss tranche, i.e. pricing further downside protection, should reflect this exposure to gap risk. Two related market events can give rise to this exposure; first, a situation where loss momentum is higher than the manager's ability to liquidate the portfolio in a timely manner. The fund is exposed to price fluctuations from the moment when the fund value first hits the barrier, at which point the manager is aware of the situation and starts the liquidation of all assets, through to the end of the liquidation process. Second, a situation where the liquidity of the underlying assets is low, to the point that an orderly liquidation of the portfolio extends in time beyond the liquidation barrier and assets can continue to lose value. Whereas highly liquid assets (e.g. most stocks, futures or certain bonds) can be liquidated within seconds or minutes to minimize the gap risk, illiquid assets (e.g. real estate or certain OTC products) take time to liquidate and are exposed to market risk over a longer period of time. As the manager only covers losses up to the barrier in both cases, the investor is exposed to the gap risk and losses can exceed the first tranche. 

We analyze a contract in which an investor pays an upfront premium to a reinsurance party and in exchange gains full portfolio protection net of fees. 
The contribution of this paper is a method of computing the premium for the reinsurance as a function of the liquidity of the underlying portfolio. Assuming risk-neutral 
valuation and a geometric Brownian motion for the value of the underlying portfolio, we derive a closed-form expression for the premium.
In addition to the analytical approach, we conduct a numerical 
simulation using a Markov-switching model as an alternative. Furthermore, we provide some sensitivity analysis with respect to certain market parameters and variations of the initial setting.

The remainder of the paper is structured as follows. Section two elaborates on established first-loss concepts and introduces our modifications. In section three, we consider the analytical model using a geometric Brownian motion and derive a closed-form solution. The fourth section contains a 
simulation in a Markov-Switching model. In section five, we present numerical results. Section six is dedicated to backtesting the introduced models. Section seven concludes. We wish to express our gratitude to Michael Ege for some very useful insights on the business aspects of this paper as well as for interesting comments of earlier drafts

\section{Hedge fund fee structures}

\subsection{Established first-loss framework}
First-loss schemes were established several years ago but have found rising interest in recent years\footnote{https://www.marketwatch.com/story/more-hedge-funds-lured-to-new-source-of-capital-2011-05-23?pagenumber=1 for an article on Dow Jones \& Company's MarketWatch about rising interest in first-loss schemes in recent years from May 23, 2011 and https://www.bloomberg.com/news/articles/2018-04-23/paulson-levers-up-bets-seeking-big-fees-while-risking-his-money for a Bloomberg article on recent first-loss related news from April 23, 2018.}. Especially in the aftermath of the financial crisis, managers lacked investor confidence. The first-loss principles not only signal a manager's dedication, as his own funds are at stake, but also allow smaller funds and those without a long track record to attract new investors and quickly grow assets in the early seeding phase. However, the first-loss scheme comes with a price. The manager provides downside protection by covering the first tranche of portfolio losses in return for a higher participation rate in the event of a positive fund performance. Whereas in traditional fee structures the manager receives a flat 2\% management fee and a 20\% performance-dependent fee on net profits (known as `2 and 20'), a first-loss structure might allow a manager to raise the performance fee to 50\% by insuring losses up to 10\% of the initial investment.

There exists a variety of first-loss variations. In this paper, we are closely following the framework of \textcite{secosharedloss}. The fund value $X_t$ consists of two	parts, the manager's part $V_M(t)$ and the investor's part $V_I(t)$, where $X_t = V_I(t) + V_M(t)$. The payoffs at the terminal time $T$ are defined as follows:
\begin{align*}
\begin{array}{lll}
	V_I(T) &= \left\lbrace
			\begin{array}{l}
			X_T - m_M X_0 - \alpha_M (X_T - m_m X_0 - X_0)^+ \\
			(1 - m_M) X_0 \\	
			X_T + (c_M - m_M) X_0
			\end{array}	
			\right.
				&\begin{array}{l}
				when \, X_T \geq X_0\\
				when \, (1 - c_M) X_0 \leq X_T \leq X_0\\	
				when \, X_T \leq (1-c_M)X_0
			\end{array}
\\
\\
	V_M(T) &= \left\lbrace
			\begin{array}{l}
			m_M X_0 + \alpha_M (X_T - m_m X_0 - X_0)^+ \\
			m_M X_0 - (X_0 - X_T) \\	
			m_M X_0 - c_M X_0 
			\end{array}	
			\right.
				&\begin{array}{l}
				when \, X_T \geq X_0\\
				when \, (1 - c_M) X_0 \leq X_T \leq X_0\\	
				when \, X_T \leq (1-c_M)X_0
			\end{array}
\end{array}
\end{align*}
or, in more condensed form:
\begin{align}
V_I(T) &= X_T - m_M X_0 - \alpha_M [X_T - m_m X_0 - X_0]^+ + [X_0 - X_T]^+ - [(1-c_M)X_0 - X_T]^+ \label{payoff I old}\\
V_M(T) &= m_M X_0 + \alpha_M [X_T - m_m X_0 - X_0]^+ - [X_0 - X_T]^+ + [(1-c_M)X_0 - X_T]^+ \label{payoff M old}
\end{align}
where $X_0$ is the initial investment, $\alpha_M$ is the performance dependent fee, $m_M$ is the flat management fee and $c_M$ the managerial deposit, i.e. $(1-c_M)X_0$ is the lower level of the first tranche insurance. In other words, the manager compensates the investor in the event of negative performance with up to $c_M$ of the initial investment. 

The payoff functions form the base case of our derivative-pricing framework. The investor holds a long position representing her investment, i.e. the fund's assets net fees ($X_T - m_M X_0$), a short position in a call option on the fund's assets with strike $m_M X_0 + X_0$ representing potential performance dependent fees, a long position in a put option on the fund's assets with strike $X_0$ representing first-loss insurance by the manager and a short position in a put option on the fund's assets with strike $(1-c_M)X_0$ representing the cap on the manager's deposit, i.e. the manager's limited liability in the event of a negative fund performance beyond the first-loss protection. The manager holds the respective counterparts, i.e. a long call option, a long put option and a short put option.

\subsection{Modified framework: Reinsurance}
Theoretically, by introducing a liquidation barrier at the level of the lower end of the managerial deposit (or first tranche of losses), i.e. at $K = (1 - c_M)X_0$ the investor gains full portfolio protection. In this setting, the liquidation is triggered as soon as the portfolio value breaches the barrier, in which case the manager compensates for all resulting losses. As noted 
above, however, in practice the investor is exposed to gap risk owing to potential delays when liquidating the portfolio, e.g. due to market conditions or the illiquid nature of the 
instruments in the investment portfolio.

Consider the situation in which the investor pays a premium to a third party reinsurer, in return for protection from losses not covered by the manager. Denote the value of the 
reinsurer's position by $V_{R}(t)$. 
The fund value now follows $X_t = V_I(t) + V_M(t) + V_R(t)$, where the payoff functions are defined as follows:

\begin{footnotesize}
\begin{align*}
\begin{array}{lll}
	V_I(T) &= \left\lbrace
			\begin{array}{l}
			X_T - (m_M + m_R \cdot e^{rT}) X_0 - \alpha_M (X_T - m_M X_0 - X_0)^+ \\
			(1 - m_M - m_R \cdot e^{rT}) X_0 
			\end{array}	
			\right.
				&\begin{array}{l}
				when \, X_T \geq X_0\\
				when \, X_T \leq X_0
			\end{array}
\\
\\
	V_M(T) &= \left\lbrace
			\begin{array}{l}
			m_M X_0 + \alpha_M (X_T - m_M X_0 - X_0)^+ \\
			m_M X_0 - (X_0 - X_T) \\	
			m_M X_0 - c_M X_0 
			\end{array}	
			\right.
				&\begin{array}{l}
				when \, X_T \geq X_0\\
				when \, (1 - c_M) X_0 \leq X_T \leq X_0\\	
				when \, X_T \leq (1-c_M)X_0
			\end{array}
\\
\\
	V_R(T) &= \left\lbrace
			\begin{array}{l}
			m_R \cdot e^{rT} \cdot X_0 \\	
			m_R \cdot e^{rT} \cdot X_0 - ((1 - c_M) X_0 - X_T)
			\end{array}	
			\right.
				&\begin{array}{l}
				when \,X_T \geq (1-c_M)X_0\\
				when \, X_T \leq (1-c_M)X_0
			\end{array}
\end{array}
\end{align*}
\end{footnotesize}
where $X_0$ is the initial investment, $\alpha_M$ is the performance dependent fee, $m_M$ is the flat management fee and $c_M$ the managerial deposit, i.e. $(1-c_M)X_0$ is the lower level of the first tranche insurance, $m_R$ is the upfront premium for the full portfolio insurance and $r$ is the risk-free interest rate. Written more compactly:
\begin{align}
V_I(T) &= X_T - (m_M + m_R \cdot e^{rT}) X_0 - \alpha_M[X_T - m_M X_0 - X_0]^+ + [X_0 - X_T]^+ \label{payoff I new}\\
V_M(T) &= m_M X_0 + \alpha_M [X_T - m_M X_0 - X_0]^+ - [X_0 - X_T]^+ + [(1-c_M)X_0 - X_T]^+ \label{payoff M new}\\
V_R(T) &= m_R \cdot e^{rT} \cdot X_0 - [(1 - c_M) X_0 - X_T]^+ \label{payoff R new}
\end{align}
Note we assume that the investor cannot redeem her investment during the investment horizon, e.g. due to an imposed lock-up period even though it might be optimal to prematurely withdraw from the arrangement from the point of view of the investor (see \textcite{saunders2016analysis} and \textcite{meng2016optimal}, who analyze optimal stopping times). The payoff functions (\ref{payoff I new}) to (\ref{payoff R new}) extend the functions in (\ref{payoff I old}) and (\ref{payoff M old}). The short position representing the cap on the manager's deposit is transferred from the investor to the reinsurer, i.e. the investor gains full portfolio protection net of fees. However, the investor is subject to another future liability representing the premium ($m_R \cdot e^{rT} \cdot X_0$). Despite the full portfolio protection, the investor is now facing a certain credit risk, as she now relies on the financial solvency of the reinsurance in the event of a negative fund performance. We assume no default and thus neglect the credit risk. Note that, from the manager's perspective, nothing changes, as (\ref{payoff M old})=(\ref{payoff M new}). The transaction only involves the investor and reinsurer.

As this paper focuses on the upfront premium paid by the investor, we will be placing the emphasis on the payoff of the reinsurance in (\ref{payoff R new}) hereafter. Note that the payoff function in (\ref{payoff R new}) is defined assuming no liquidation barrier, i.e. computed using the terminal fund value at the end of the investment horizon $T$. In order to incorporate the aforementioned gap risk and thus reflect a fund's level of liquidity, we need some alterations of the payoff function of the reinsurance stated in (\ref{payoff R new}). We set a liquidation barrier at the lower end of the first tranche, i.e. at $K=(1-c_M) X_0$. If the fund value breaches this barrier at any time $\tau \in (0,T)$, all operations stop and the remaining assets are fully liquidated. 
Reflecting a fund's underlying level of liquidity, we introduce the parameter $\Theta$. As a measure of liquidity, $\Theta$ is defined as the time a fund needs to liquidate its assets. The payoff function is defined as follows:
\begin{align}
V_R(\tau + \Theta) &= m_R \cdot e^{r (\tau + \Theta)} X_0 - [K - X_{\tau + \Theta}]^+ \; &&\tau \, \in \, (0,T) \label{payoff R final}
\end{align}
where $\tau = \inf\{ t>0 ; X_{t} \leq K\}$.
In the event, that the fund value does not breach the liquidation barrier, we have $[K-X]^+ \overset{X>K}{=} 0$, i.e. the reinsurer receives and keeps the full premium. 
The benchmark value that we use is $\Theta = \frac{1}{252}$, i.e. one day until full liquidation of the fund. In practice, the liquidation process of a fund heavily depends on the composition of its financial assets and the liquidation time is different for each asset. 
In the following sections we derive the price of the upfront premium depending on the fund's liquidity. 

\section{Analytic Valuation using Geometric Brownian Motion}
\label{Analytic Valuation using Geometric Brownian Motion}

Let the hedge fund performance $X=\{X_t\}_{t\in [0,T]}$ follow a geometric Brownian motion. Under the risk-neutral measure 
$\mathbb{Q}$, we have
\begin{align}
dX_t &= X_t \cdot (rdt+\sigma d\tilde{W}_t) \nonumber\\
\Rightarrow X_t &= X_0 \cdot e^{(r-\frac{1}{2}\sigma^2)t+\sigma \tilde{W}_t} \label{hf process}
\end{align}
where $r$ is the risk-free rate, $\sigma > 0$ is the fund's volatility and $\tilde{W}_t$ is a $\mathbb{Q}$-Brownian motion.
As above, let $\tau := \inf\{0 < t \leq T:X_t = K\}$ be the stopping time at which the fund value hits the barrier $K = (1-c_m)\cdot X_0$. At time $\tau$, the fund is shut down and all its assets are liquidated. Using (\ref{hf process}), we have 
\begin{align}
\tau &:= \inf\{0 < t \leq T:X_t = K\} = \inf\{0 < t \leq T:Y_t = a\} \label{hitting time}
\end{align} 
where $Y_t = \underbrace{(r-\frac{1}{2}\sigma^2)}_{\mu}t+\sigma \tilde{W}_t$ and $a = \ln(\frac{K}{X_0})$. 
For fixed $a$, the first hitting time $\tau$ in (\ref{hitting time}) follows an inverse Gaussian distribution (see \textcite{kwok2008mathematical}) 
$\tau \sim IG(\mu^*,\lambda^*)$ where $\mu^*=\frac{a}{\mu}$, $\lambda^*=\frac{a^2}{\sigma^2}$ and the density function $f$ is:
\begin{align}
f(\tau,\mu^*,\lambda^*) &= \sqrt{\frac{\lambda^*}{2 \pi \tau^3}} \exp(-\frac{\lambda^*(\tau-\mu^*)^2}{2 (\mu^*)^2 \tau}) \nonumber \\
\Leftrightarrow f(\tau,a,\mu) &= \frac{|a|}{\sigma \sqrt{2 \pi \tau^3}} \exp(-\frac{(a-\mu \tau)^2}{2 \sigma^2 \tau}) \label{inverse gaussian distribution}
\end{align}


We derive the value of the upfront premium $m_R$ as a function of a fund's liquidity by using the fair value approach stated in \textcite{secosharedloss}, where the fair value for the premium is obtained by setting the present value of the expected payoff equal to the initial investment, i.e. $V_R(0) = 0$. In our setting we have:
\begin{align}
V_R(0) &= \mathbb{E}_{\mathbb{Q}}[e^{-r(\tau + \Theta)} V_R(\tau + \Theta)] = 0 \label{fair value approach}
\end{align}

The payoff made by the reinsurer is that of a put option issued at time $\tau$ with strike $K = (1-c_M) X_0$ and exercise time $\tau + \Theta$.
For simplicity and without loss of generality, we set the initial investment to $X_0=1$. From (\ref{payoff R final}), the terminal value of the reinsurer at time $\tau + \Theta$ is thus:
\begin{align}
V_R(\tau + \Theta) = m_R \cdot e^{r (\tau + \Theta)} X_0 - [K - X_{\tau + \Theta}]^+ 
= m_R \cdot e^{r (\tau + \Theta)} - [K - X_{\tau + \Theta}]^+
\end{align}
At time $\tau$, the time of issuance of the put option, the value of the reinsurer's position is:
\begin{align}
V_R(\tau) &= \mathbb{E}_{\mathbb{Q}}[e^{-r\Theta} \cdot V_R(\tau + \Theta)|\mathcal{F}_{\tau}]= m_R \cdot e^{r\tau} - V_1(\Theta)  \nonumber
\end{align}

By definition $K = X_{\tau}$, as the option is issued when the fund breaches the barrier. Thus, $V_1(\Theta)$ follows the well-known Black-Scholes option formula:
\begin{align}
V_1(\Theta) = \mathbb{E}_{\mathbb{Q}}[e^{-r \Theta} [K - X_{\tau + \Theta}]^+|\mathcal{F}_{\tau}] = K (e^{-r\Theta} \cdot N(-d_2) - N(-d_1)) \nonumber\\
\mbox{with\;\;} \, d_1 = \frac{(r+\frac{1}{2}\sigma^2)\sqrt{\Theta}}{\sigma}, \; 
d_2 = d_1 - \sigma\sqrt{\Theta} \label{V1}
\end{align}
where $N(\cdot)$ is the standard normal cumulative distribution function.
Note that $V_1(\Theta)$ does not depend on $\tau$, but solely on $\Theta$, i.e. $V_1(\Theta)$ is a function of liquidity. At time $t=0$, the initial value of the reinsurer is:
\begin{align}
V_R(0) = \mathbb{E}_{\mathbb{Q}}[e^{-r\tau} \cdot V_R(\tau)|\mathcal{F}_0] = m_R - V_1(\Theta) \cdot 
\mathbb{E}_{\mathbb{Q}}[e^{-r\tau}|\mathcal{F}_0] 
= m_R - V_1(\Theta) \cdot V_2(T) \label{value at zero}
\end{align}
where
\begin{align}
V_2(T) = \mathbb{E}_{\mathbb{Q}}[e^{-r\tau}|\mathcal{F}_0] = \int_{0}^{T} e^{-rt} f(t)\, dt \label{integral}
\end{align}
The above integral can be evaluated analytically (see \textcite{kwok2008mathematical}):
\begin{align}
V_2(T) = \left(\frac{K}{X_0}\right)^{\alpha_{+}} N\left(\delta \frac{\ln(\frac{K}{X_0})+\beta T)}{\sigma \sqrt{T}}\right) + \left(\frac{K}{X_0}\right)^{\alpha_{-}} N\left(\delta \frac{\ln(\frac{K}{X_0})-\beta T)}{\sigma \sqrt{T}}\right) \nonumber \\
\mbox{with} \;\; \beta = \sqrt{(r-\frac{\sigma^2}{2})^2+2r\sigma^2}, \alpha_{\pm} = \frac{r-\frac{\sigma^2}{2}\pm\beta}{\sigma^2} \;\; \mbox{and} \;\, \delta = \mbox{sign}\left(\ln(\frac{X_0}{K})\right).\label{V2}
\end{align} 
We thus obtain:
\begin{equation} 
m_{R} = V_1(\Theta) \cdot V_2(T) \label{premium}
\end{equation}

The premium is composed of two parts: $V_1(\Theta)$ represents the value of the issued option and $V_2(T)$ incorporates the (discounted) probability of breaching the liquidation barrier throughout the entire investment horizon $T$. 

Note, some practitioners might prefer results for a model which is fitted to historical data. Hence, we replace the risk-free drift with an empirical drift $b$, which implies that $\mbox{log}(\frac{X_{t+\Delta}}{X_t})\sim N((b-\frac{1}{2}\sigma^2)\Delta,\sigma^2 \Delta)=N(\mu_{emp}\Delta,\sigma_{emp}^2\Delta)$ for a daily setting with $\Delta = \frac{1}{252}$ as log-returns are normally distributed. Best estimates for drift and volatility are $\hat{\sigma}=\sigma_{emp}$ and $\hat{b}=\mu_{emp}+\frac{1}{2}\sigma_{emp}^2$, where $\mu_{emp} = 252 \cdot \frac{1}{N}\sum\limits_{n=1}^{N} R_n$ and $\sigma_{emp} = \sqrt{252 \cdot \frac{1}{N-1}\sum\limits_{n=1}^{N} (R_n - \mu_{emp})}$ with daily log-return $R_n$. The empirical yearly returns $\mu_{emp}$ and volatility $\sigma_{emp}$ for selected hedge fund indices (see details in section \ref{section Valuation in a Markov-switching framework}) can be found in table \ref{table HFRX}\footnote{Estimating drift rates is notoriously difficult and nearly impossible to do in practice (see \textcite{merton1980estimating} for the original paper on the problem). We follow a simple approach and use mean and standard deviation of log-returns.}. We would like to point out that we are not pricing the contract using the risk-neutral measure $\mathbb{Q}$ in the following. The crucial assumptions needed to justify the replication arguments that underlie risk-neutral valuation might not hold in the hedge fund context, e.g. as we specifically allow illiquid asset classes, the arbitrage-free dynamic hedging strategy underlying the classic Black-Scholes framework is hardly attainable. Therefore, we evaluate the present value of the expected payoff (represented by $V_1^{\mathbb{P}}(\Theta)$ in equation (\ref{V_1_rw})). Investors might be interested in this value, which is still available in closed-form.
For equation (\ref{V1}) we have:
\begin{align}
V_1^{\mathbb{P}}(\Theta) = \mathbb{E}_{\mathbb{P}}[e^{-r \Theta} [K - X_{\tau + \Theta}]^+|\mathcal{F}_{\tau}] = K (e^{-r\Theta} \cdot N(-d_2) - e^{(b-r)\Theta} \cdot N(-d_1)) \nonumber\\
\mbox{with\;\;} \, d_1 = \frac{(b+\frac{\sigma^2}{2})\sqrt{\Theta}}{\sigma}, \; 
d_2 = d_1 - \sigma\sqrt{\Theta}\label{V_1_rw}
\end{align}
and for equation (\ref{V2}) we have:
\begin{align}
V_2^{\mathbb{P}}(T) = \left(\frac{K}{X_0}\right)^{\alpha_{+}} N\left(\delta \frac{\ln(\frac{K}{X_0})+\beta T)}{\sigma \sqrt{T}}\right) + \left(\frac{K}{X_0}\right)^{\alpha_{-}} N\left(\delta \frac{\ln(\frac{K}{X_0})-\beta T)}{\sigma \sqrt{T}}\right) \nonumber \\
\mbox{with} \;\; \beta = \sqrt{(b-\frac{\sigma^2}{2})^2+2r\sigma^2}, \alpha_{\pm} = \frac{(b-\frac{\sigma^2}{2})\pm\beta}{\sigma^2} \;\; \mbox{and} \;\, \delta = \mbox{sign}\left(\ln(\frac{X_0}{K})\right).
\end{align}
We thus obtain:
\begin{align}
m_{R}^{\mathbb{P}} = V_1^{\mathbb{P}}(\Theta) \cdot V_2^{\mathbb{P}}(T)
\end{align}

\section{Valuation in a Markov-switching framework}
\label{section Valuation in a Markov-switching framework}
We consider a Markov-switching model and evaluate the payoff of the reinsurance 
to compute the premium $m_R$. Markov-switching or regime-switching models were first introduced by \textcite{hamilton1989new} and have since found regular application in the financial world. These models create returns similar to actual hedge fund returns and show typical hedge fund characteristics, i.e. skewness, volatility clusters and fat tails. We assume a two state Markov-switching model with state space $\mathbb{S} = \{1,2\}$ for the continuous-time Markov chain $\varepsilon(t)$, where state 1 represents the `normal' market and state 2 the `stressed' market. $\varepsilon(t)$ is generated by the intensity matrix
\begin{align*}
	Q = \left[\begin{array}{cc}
		-\lambda_1 & \lambda_1 \\
		\lambda_2 & -\lambda_2
	\end{array}\right],
\end{align*}
where $\lambda_1$ and $\lambda_2$ are the transition rates for switching states. The fund value satisfies the following stochastic differential equation:
\begin{align*}
dX_t = X_t \cdot (\mu_{\varepsilon(t)}dt + \sigma_{\varepsilon(t)} dW_t) \, , \, X_0 > 0
\end{align*}
where $\varepsilon(t) \in \{1,2\}$ and $\mu_{\varepsilon(t)}$ and $\sigma_{\varepsilon(t)}$ represent the respective state's return rate and volatility. 
In order to use the risk-neutral martingale pricing methodology, we need an equivalent martingale measure $\mathbb{Q}$. We follow the approach in \textcite{henriksen2011pricing} by using the Girsanov theorem and assume unchanged transition probabilities under the risk-neutral measure $\mathbb{Q}$ (see \textcite{bollen1998valuing} and \textcite{hardy2001regime} for the same approach). The resulting price process for constant $r$ is given by:
\begin{align*}
X_t = X_0 \cdot \exp\left\lbrace\int_0^t r-\frac{1}{2}\sigma_{\varepsilon(u)}^2)du + \int_0^t \sigma_{\varepsilon(u)} d\tilde{W}_u\right\rbrace \, , \, X_0 > 0.
\end{align*}
For another approach see \textcite{elliott2005option}, where a regime switching Esscher transform is applied to obtain an equivalent measure. Note, the Markov-switching modeled market is incomplete, as there is an additional stochastic dimension in form of the regime switching and a unique $\mathbb{Q}$ measure cannot be determined (for a proof see for example \textcite{elliott2007pricing}).

When performing simulations, we implement the following discretized version of the regime-switching Markov process on an equidistant grid:
\begin{align*}
X_{t_i} &= X_0 \cdot e^{R_{t_i}}, \, \forall i \in \{0,...,k\}, \, X_0 > 0 \\
R_{t_{i}} &= R_{t_{i-1}} + \left((r-\frac{1}{2}\sigma_{\tilde{\varepsilon}(t_i)}^2)\Delta + \sigma_{\tilde{\varepsilon}(t_i)}\sqrt{\Delta} \cdot \eta_{t_i}\right), \, R_{t_0} = 0
\end{align*}
where $0 = t_0 < t_1 < ... < t_k = T$, $\Delta \equiv \Delta t_j := t_{j+1} - t_j = T/k$, $\eta_{t_i}$ are i.i.d. standard normal random variables and $\tilde{\varepsilon}(t)$ is a discretization of the continuous-state Markov-chain $\varepsilon(t)$ with transition matrix:
\begin{align*}
	P = \left[\begin{array}{cc}
		(1-p) & p \\
		q & (1-q)
	\end{array}\right],
\end{align*}
where p and q are the transition probabilities for switching from one state to the other:  
\begin{gather}
p = \mathbb{P}(\tilde{\varepsilon}(t_{i+1}) = 1 | \tilde{\varepsilon}(t_i) = 2)\nonumber \\
q = \mathbb{P}(\tilde{\varepsilon}(t_{i+1}) = 2 | \tilde{\varepsilon}(t_i) = 1)\nonumber 
\end{gather} 
The time-homogeneous Markov chain $\tilde{\varepsilon}$ has stationary distribution $\pi_1 = \frac{q}{q+p}$ and $\pi_2 = \frac{p}{q+p}$. We  assume daily time steps and consider one year to be made up of 252 trading days, i.e. $\Delta = \frac{1}{252}$.

For the numerical approach, we simulate $N = 100,000$ paths of the discretized Markov-switching process\footnote{Simulations were conducted using MatLab and R.}. For each sample path $X_t^n$, $n = 1,...,N$, the reinsurer's value at time $t = \tau + \Theta$ is given by:
\begin{align}
V_R^n(\tau^n + \Theta) &= m_R^n \cdot e^{r (\tau^n + \Theta)} X_0 - [K - X^n_{\tau^n + \Theta}]^+.
\end{align}
At time $t=0$ the reinsurer's value is thus:
\begin{align}
V_R^n(0) &= m_R^n X_0 - e^{-r (\tau^n + \Theta)} [K - X^n_{\tau^n + \Theta}]^+. \label{payoff ms theta}
\end{align}
Similar to the analytical valuation, we use the fair value approach stated in \textcite{secosharedloss} to derive the upfront premium $m_R$ and set the present value of the expected payoff equal to zero such that:
\begin{equation} 
\frac{1}{N} \sum_{n=1}^{N} V_{R}^{n}(0) = 0
\end{equation} 
i.e. using equation (\ref{payoff ms theta}) and $X_0 = 1$ we find
\begin{equation} 
\frac{1}{N} \sum_{n=1}^{N} (m_{R} X_{0} - e^{-r(\tau^{n}+\Theta)}[K-X_{\tau^{n}+\Theta}^{n}]^{+}) = 0
\end{equation}
We thus obtain:
\begin{align}
m_R &= \frac{1}{N} \sum_{n=1}^{N} e^{-r(\tau^{n}+\Theta)}[K-X_{\tau^{n}+\Theta}^{n}]^{+} \label{premium ms}
\end{align}

The following illustrates the employed algorithm:
\begin{itemize}
	\item[1)] Simulate the (discretized) Markov chain $\tilde{\varepsilon}^n(t)$ for path $n$ with state space $\mathbb{S} = \{1,2\}$ for $t \in \{\frac{1}{252},\frac{2}{252},...,T$ \} in a daily setting with 252 trading days where $T = 1$, i.e. one year.
	\item[2)] Simulate the daily trajectory  $X_t^n = X_0 \cdot e^{R_t^n}$ with $R_t^n = R_{t-1}^n + \frac{1}{252}\cdot (r-\frac{1}{2}\sigma_i^2) + \sqrt{\frac{1}{252}} \cdot \sigma_i \cdot \eta$ where $R_0 = 0$, $\eta \sim N(0,1)$ and $i \in \{1,2\}$ denotes the state of the market.
	\item[3)] If $X_t^n < 0.9 X_0$ for any $0 < t < T $, set the stopping time $\tau^n = \inf\{0<t<T: X_t^n < 0.9 X_0\}$ and use $X_{\tau^n + \Theta}^n$ to evaluate the respective payoff and $\tau^n + \Theta$ to discount the payoff.
	\item[4)] Repeat step 1 to 3 a sufficient number of times (here $N = 100,000$) and apply (\ref{premium ms}) to find the expected upfront premium $m_R$.
\end{itemize}

As it is preferable to decrease an estimate's variance, we also apply the antithetic variates method in our simulation. This method exploits the symmetry of the randomly generated i.i.d. standard normal variables $\eta \sim N(0,1)$ in step 1) of our algorithm above. It holds for $\eta \sim N(0,1)$, that $\eta \overset{d}{=} -\eta$ and thus we can replace $\eta$ by $-\eta$ in step 1) without changing the trajectory's distribution. For the estimated trajectory $X^n_{antithetic} := \frac{X^n_{\eta} + X^n_{-\eta}}{2}$ it holds, that $\mbox{var}(X^n_{antithetic}) \leq \mbox{var}(X^n)$ (see \textcite{kroese2013handbook} for details).

Again, some practitioners might be interested in the actual discounted expected payoff of their investment. Analogously to the previous section (GBM model), we replace the risk-free drift with empirical yearly returns $\mu_1$,$\mu_2$ and volatility $\sigma_1$,$\sigma_2$ for selected hedge fund indices (see details below), which can be found in Table \ref{table HFRX}. Note, the same arguments on the difficulty of estimating empirical drifts and risk-neutral pricing from the previous section hold in the Markov-switching model. We do not price the contract but evaluate the present value of the expected payoff.

We fit the parameters of the Markov-switching model to real world historical data. To obtain reasonable parameters and to cover most of the hedge fund universe, we analyze five main hedge fund strategies: Equity Hedge, Equity Market Global, Neutral, Macro and Merger Arbitrage. \textcite{saunders2013fund} use these strategies as proxies for all available strategies. Our primary source of data was the open source database for daily fund returns operated by Chicago based company Hedge Fund Research (HFR). The observation period started April 1, 2003 and ended December 28, 2018. Parameter values for $\mu_1$, $\mu_2$, $p$ and $q$ are estimated using the well-known Baum-Welch algorithm (\textcite{baum1970maximization}) in the R-routine `Baum-Welch' from the package `HiddenMarkov'.  As initial values for this maximum likelihood estimation, we employ a heuristic proposed in \textcite{ernst2009portfolio}. Finally, we use Viterbi's algorithm to create the most likely sequence of states (see \textcite{viterbi1967error} for details) and compute respective mean returns and average all five strategies' returns and transition probabilities. Another approach to obtain the required parameters is the estimation by moment matching (see \textcite{hocht2009fit} for details).

\begin{figure}[H]
\includegraphics[width = \textwidth]{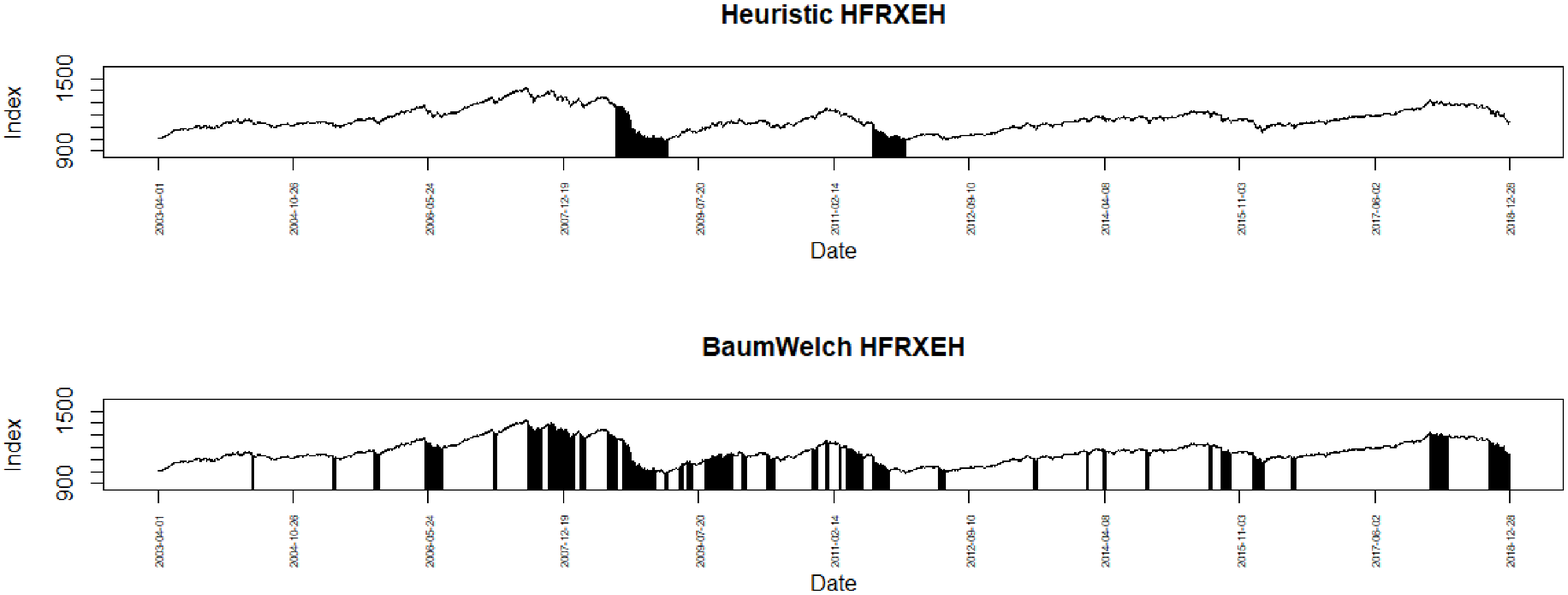}
\textit{This figure illustrates the HFRXEH Equity Hedge Index between April 1, 2003 and December 28, 2018. The top graph shows crisis periods in black using the crisis detection heuristic and the bottom graph shows crisis periods using the Baum-Welch algorithm.}
\caption{HFRXEH}
\label{HFRXEH}
\end{figure}

Figure \ref{HFRXEH} illustrates the resulting crisis periods for the exemplary HFRXEH index using the heuristic and the Baum-Welch algorithm. Table \ref{table HFRX} contains all analyzed HFRX indices and their respective numerical parameters values and historical data for a single-regime (Black-Scholes) market.

\begin{table}[H]
{\centering
\resizebox{\textwidth}{!}{\begin{tabular}{|lllcccccc|cc|}
	\hline    
    Ticker & HFRX Index Name & Strategy & $p$ [in $\%$] & $q$ [in $\%$] & $\mu_1$ [in $\%$] & $\mu_2$ [in $\%$] & $\sigma_1$ [in $\%$] & $\sigma_2$ [in $\%$] & $\mu_{emp}$ [in $\%$] & $\sigma_{emp}$ [in $\%$]\\
    \hline
    HFRXGL & Global Hedge Fund & Global & 1.67 & 7.49 & 8.54 & -32.40 & 2.57 & 6.44 & 1.08 &	3.73\\
    HFRXM & Macro\textbackslash CTA & Macro & 1.87 & 7.79 & 4.84 & -15.36 & 4.55 & 10.93 & 0.75 & 6.33\\
    HFRXEMN & EH: Equity Market Neutral & Neutral & 1.13 & 6.09 & 1.12 & -6.87 & 2.88 & 7.08 & -0.20 & 3.85\\
    HFRXEH & Equity Hedge & Equity Hedge & 2.02 & 6.19 & 11.25 & -30.36 & 4.28 & 10.12 & 0.87 & 6.35\\
    HFRXMA & ED: Merger Arbitrage & Merger Arbitrage & 2.07 & 15.70 & 5.46 & -8.27 & 2.16 & 10.17 & 3.79 & 4.02\\
    \hline
    mean  &   &   & 1.75 & 8.65 & 6.24 & -18.65 & 3.29 & 8.95 & 1.26 & 4.86 \\
    max   &   &   & 2.07 & 15.70 & 11.25 & -6.87 & 4.55 & 10.93 & 3.79 & 6.35\\
    min   &   &   & 1.13 & 6.09 & 1.12 & -32.40 & 2.16 & 6.44 & -0.20 & 3.73\\
    \hline
    \end{tabular}}}
\vspace*{0.05cm}

\textit{This table shows an overview of all HFRX indices for the period April 1, 2003 to December 28, 2018. Parameters $p,q,\mu_{1/2},\sigma_{1/2}$ were obtained from the Baum-Welch algorithm. $\mu_{emp}$ and $\sigma_{emp}$ represent empirical returns and volatility for a single-regime (Black-Scholes) market.}
\caption{Overview HFRX}     
\label{table HFRX}
\end{table}

\section{Results}

We set the initial parameters for both approaches to the following values:
\begin{multicols}{2}
	\begin{itemize}
        \item[•] $T$ = 1 (one-year investment horizon)
        \item[•] $X_0$ = 1 (initial investment)
	\end{itemize}
	\begin{itemize}
		\item[•] $K$ = (1-$c_M$)$X_0$ = 0.9 (liquidation barrier)
        \item[•] $r$ = 0.01 (risk free rate)
	\end{itemize}
\end{multicols}
In the following the unit of the liquidity measure $\Theta$ is one day, i.e. $\Theta = 1$ represents a daily liquidation window and was implemented in the model as $\frac{1}{252}$.
\subsection{Analytical results using GBM approach}

\begin{figure}[H]
\includegraphics[width = 0.5\textwidth]{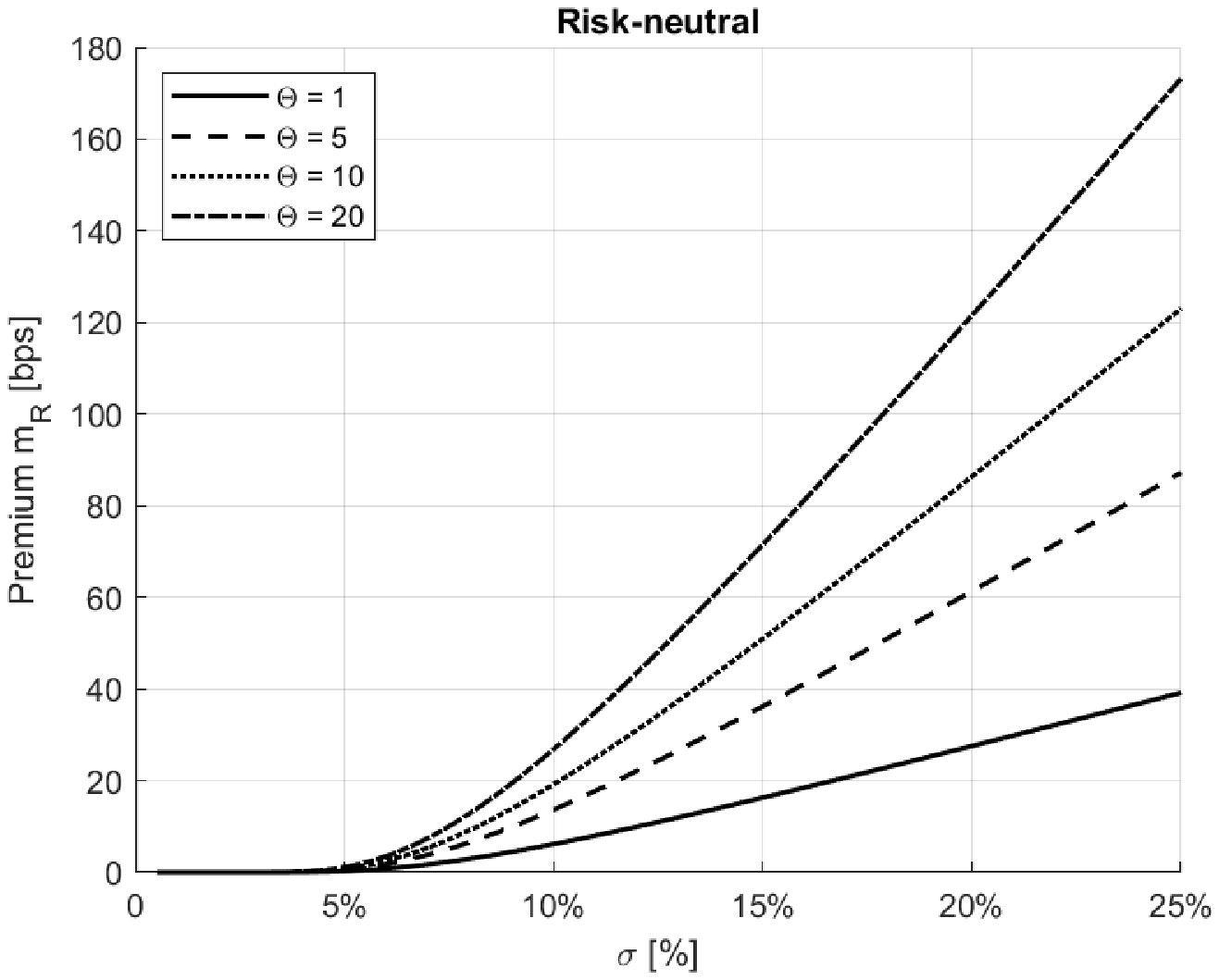}
\includegraphics[width = 0.5\textwidth]{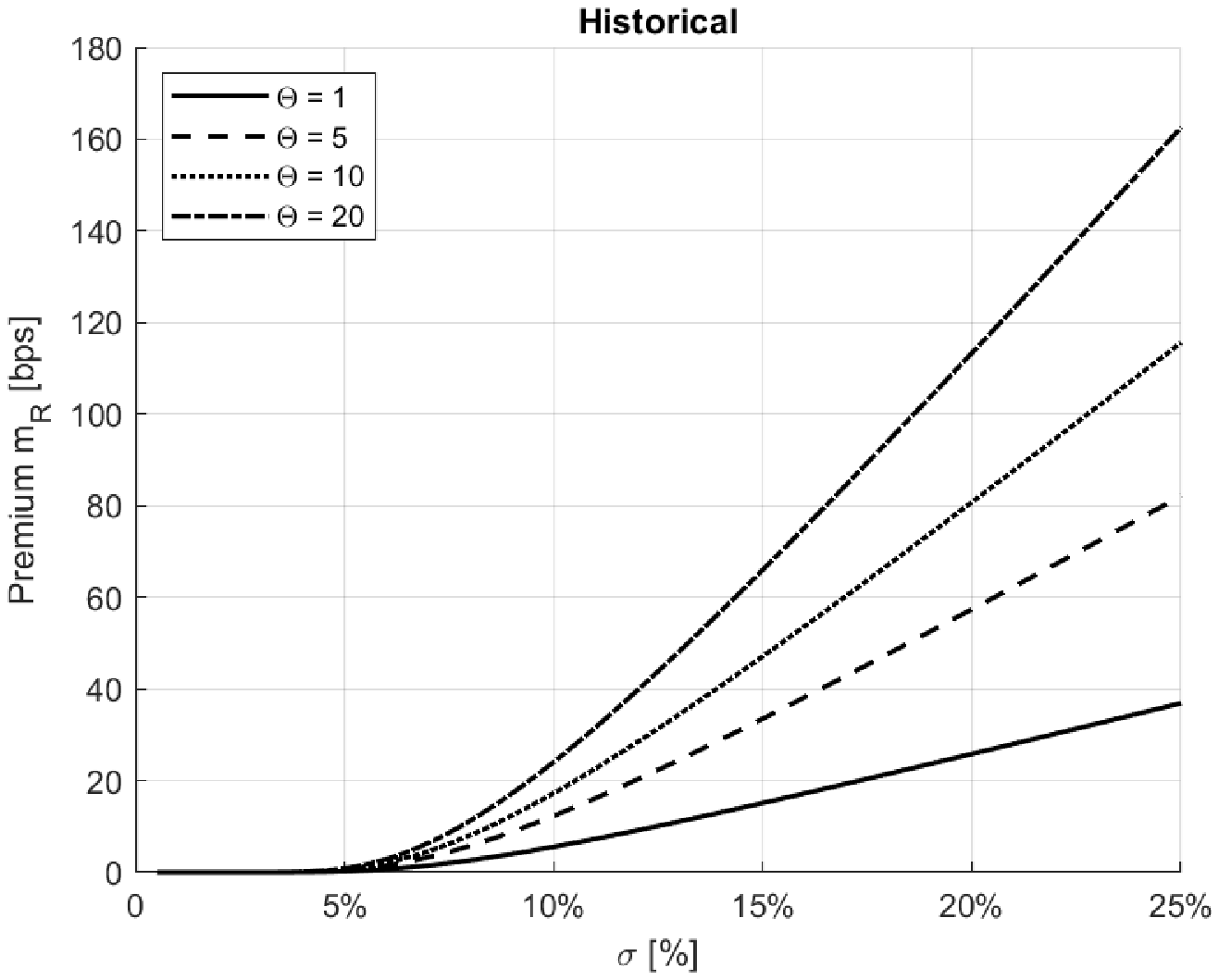}
\textit{This figure illustrates the value of the premium under the risk-neutral valuation and discounted expected payoff valuation using empirical values as a function of the volatility $\sigma$ for fixed levels of the liquidity measures $\Theta$ (in days). Parameter values are $T = 1$ (investment horizon), $X_0 = 1$ (initial investment), $c_M = 0.1$ (managerial deposit), $\mu_{emp} = 0.0126$ (empirical mean log-return) and $r = 0.01$ (risk-free rate).}
\caption{Analytical premium $m_R$ as a function of volatility}
\label{GBM (1)}
\end{figure}

\begin{figure}[H]
\includegraphics[width = 0.5\textwidth]{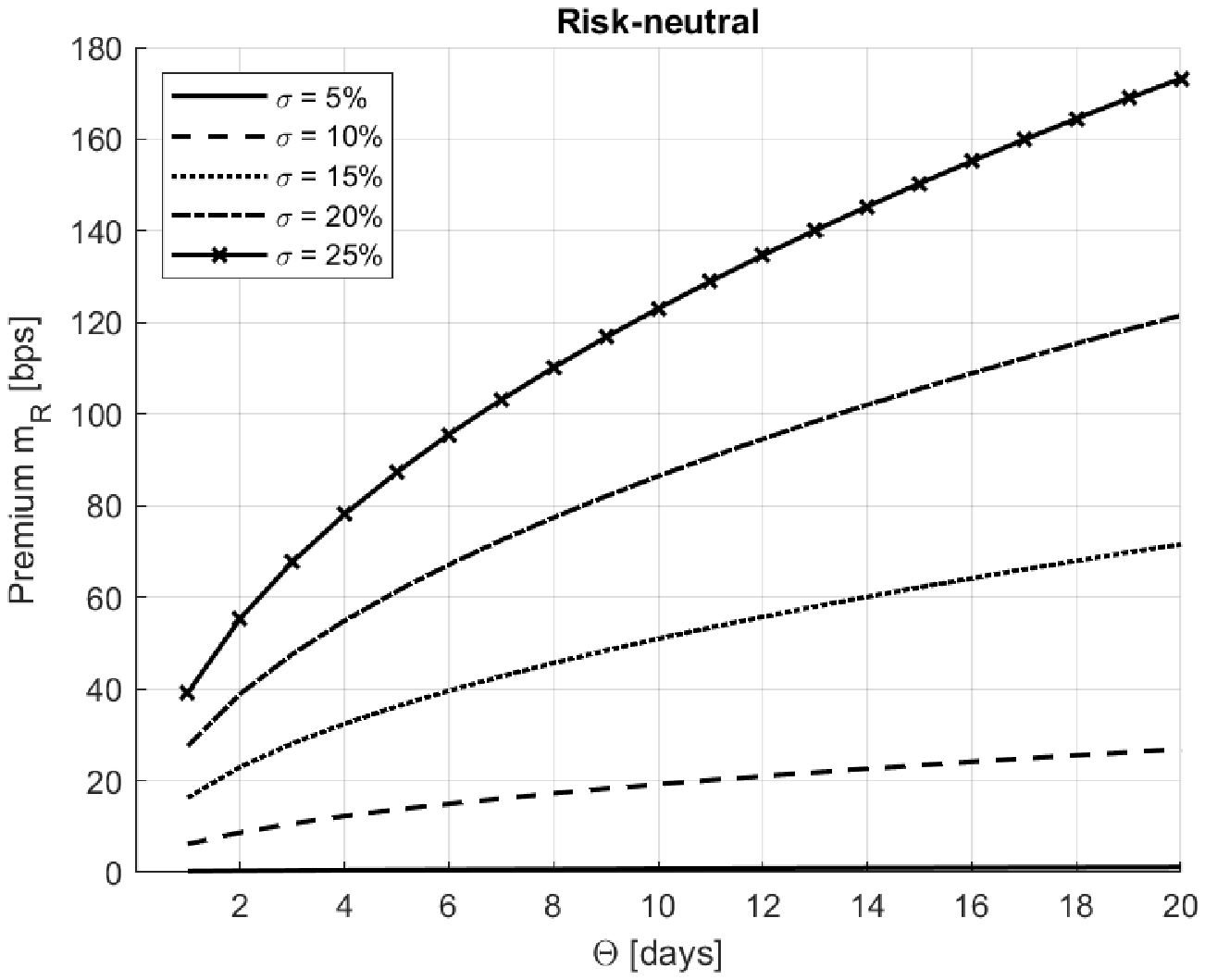}
\includegraphics[width = 0.5\textwidth]{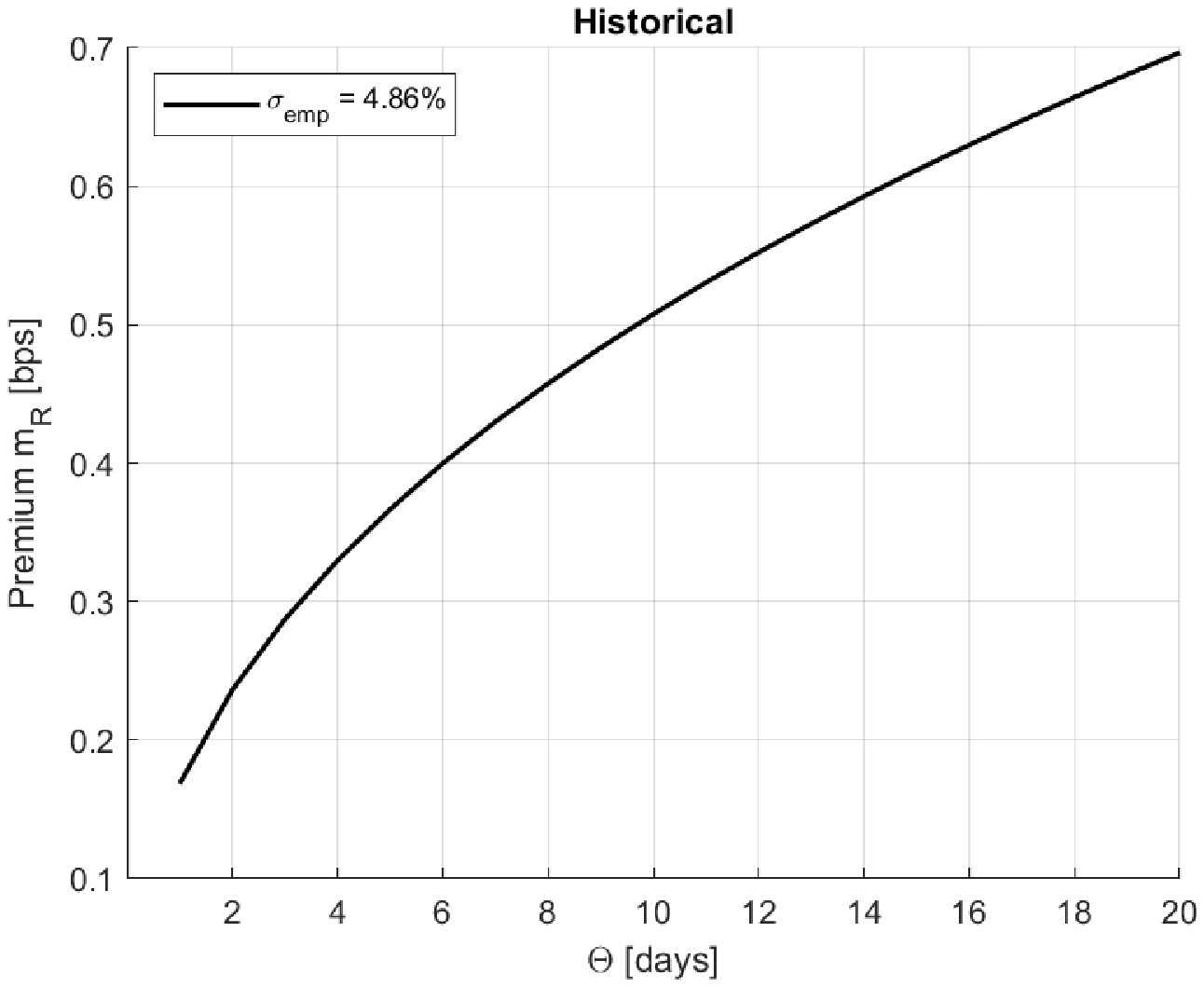}
\textit{This figure illustrates the value of the premium under the risk-neutral valuation and discounted expected payoff valuation using empirical values as a function of the liquidity measure $\Theta$ (in days) for fixed levels of volatility $\sigma$ and the empirical volatility $\sigma_{emp} = 0.0486$. Parameter values are $T = 1$ (investment horizon), $X_0 = 1$ (initial investment), $c_M = 0.1$ (managerial deposit), $\mu_{emp} = 0.0126$ (empirical mean log-return) and $r = 0.01$ (risk-free rate).}
\caption{Analytical premium $m_R$ as a function of liquidity}
\label{GBM (2)}
\end{figure}

Figures \ref{GBM (1)} and \ref{GBM (2)} show results for the analytical approach using a geometric Brownian motion framework. The premium $m_R$ is generally an increasing function of the underlying fund's volatility and liquidity. 
Figure \ref{GBM (1)} illustrates the value of premiums for fixed levels of liquidity (daily, weekly, biweekly and monthly, i.e. $\Theta \in \{1,5,10,20\}$) under the risk-neutral valuation (left graph) and discounted expected payoff valuation (right graph) using empirical values as a convex function of the volatility $\sigma$. Regardless of the liquidity window, it takes an annual volatility higher than approximately $5\%$ to observe a noticeable impact on the premium $m_R$. Assuming very liquid assets ($\Theta = 1$), we see premiums below $40bps$ even for extreme volatility ($\sigma = 25\%$). The premium using the risk-neutral valuation tends to be a bit higher than the premium computed from the discounted expected payoff valuation using empirical values (keeping all other parameters constant). Figure \ref{GBM (2)} illustrates the premium for fixed levels of volatility as a function of a fund's liquidity and the concave relationship between the premium and the liquidity. Again, for a low level of volatility ($\sigma = 5\%$), there is only a marginal impact on the premium compared to extreme levels of volatility in the risk-neutral valuation approach (left graph). The right graph illustrates the valuation of the discounted expected payoff. Even for a one-month liquidation window ($\Theta = 20$) the premium is at a reasonable value of $m_R \approx 0.7bps$.

\subsection{Numerical results using Markov-switching approach}

\begin{figure}[H]
\includegraphics[width = 0.5\textwidth]{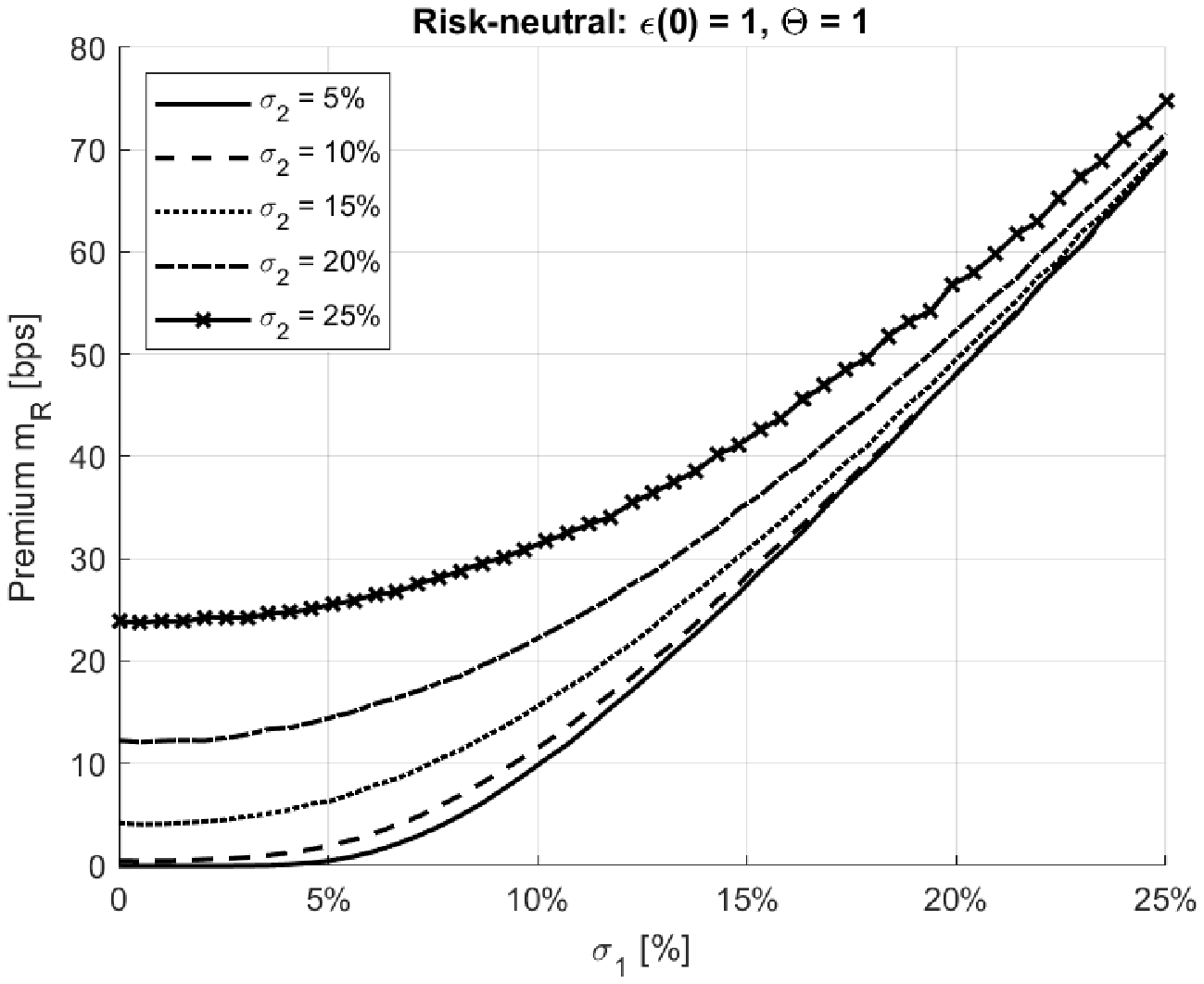}
\includegraphics[width = 0.5\textwidth]{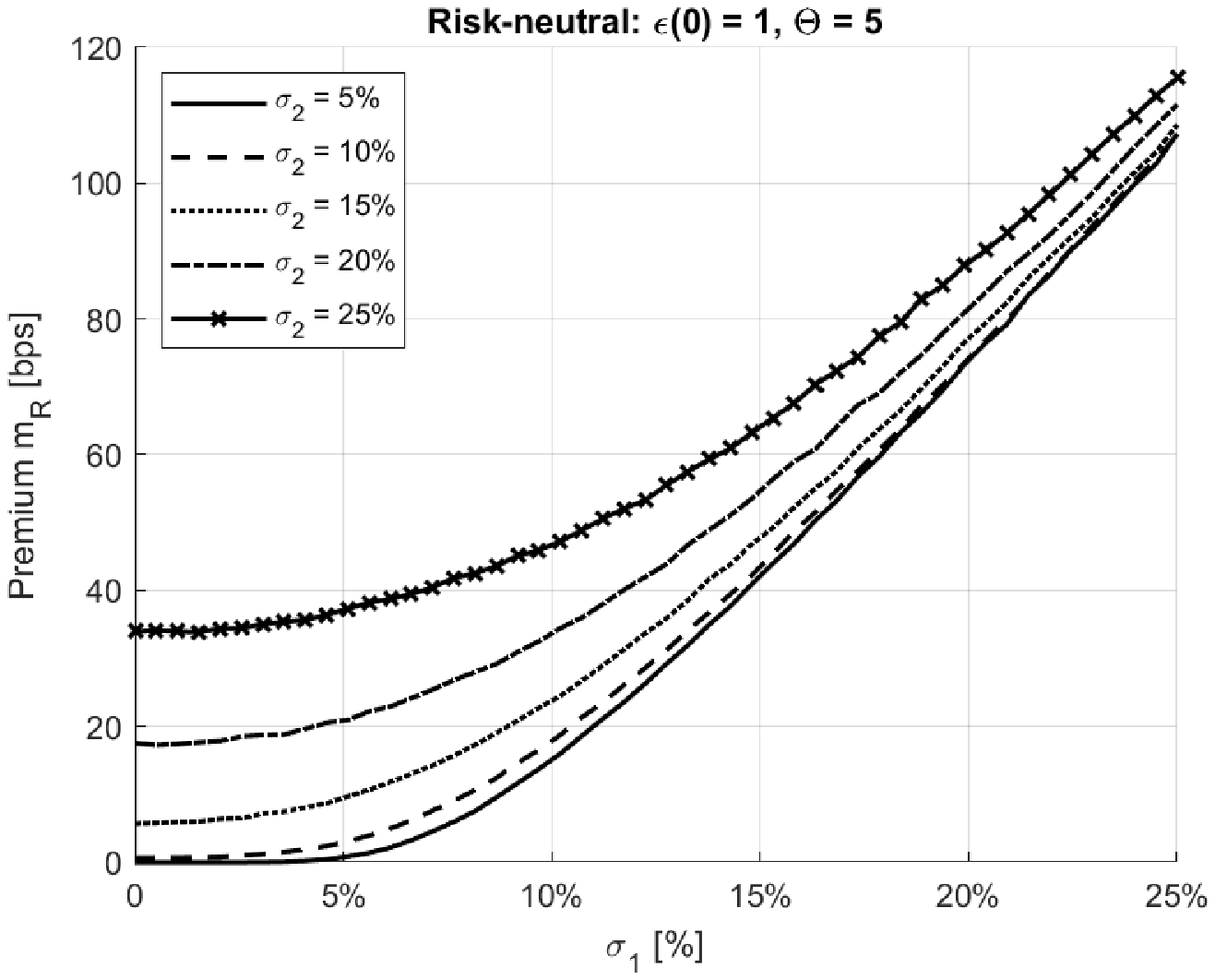} \\
\includegraphics[width = 0.5\textwidth]{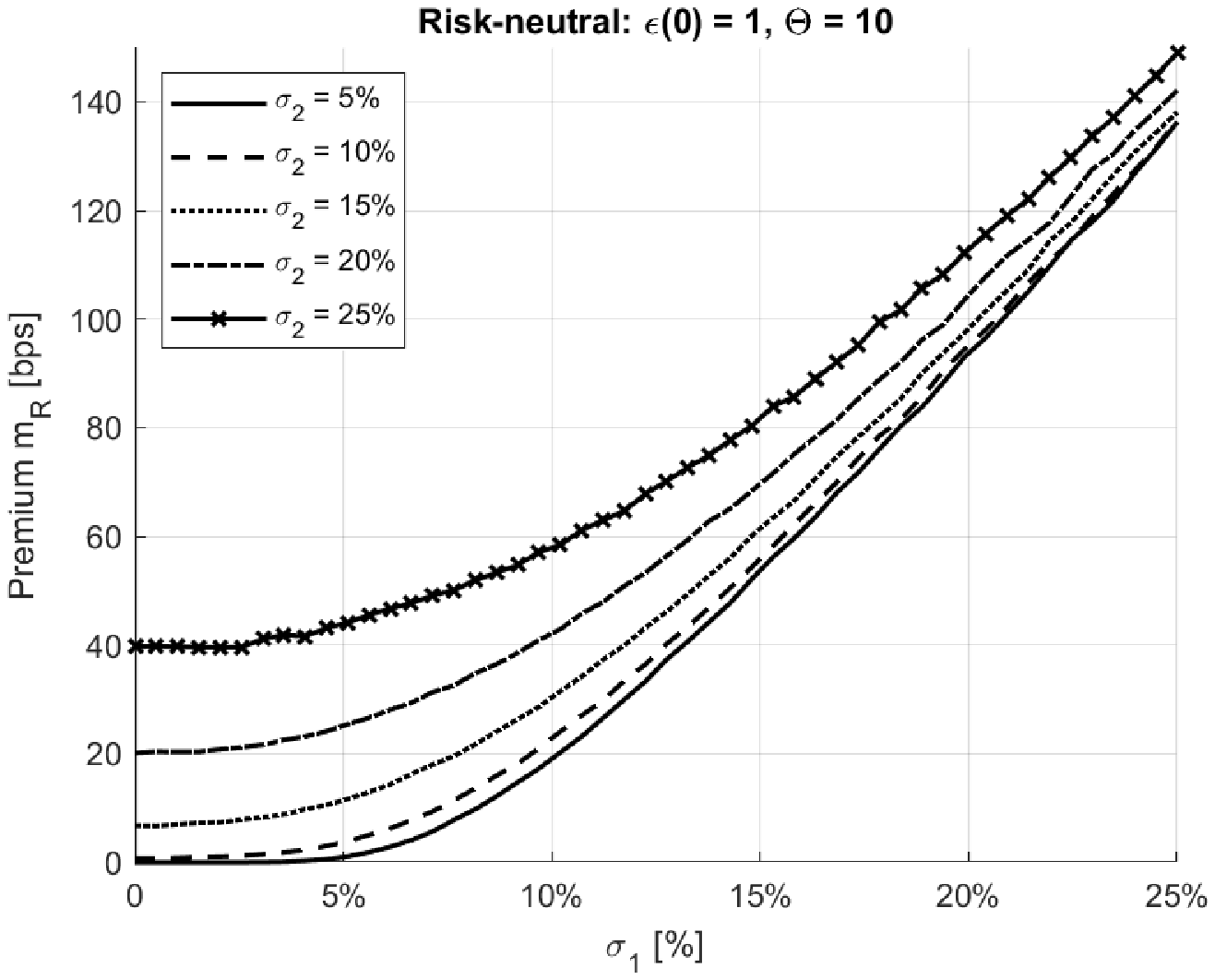}
\includegraphics[width = 0.5\textwidth]{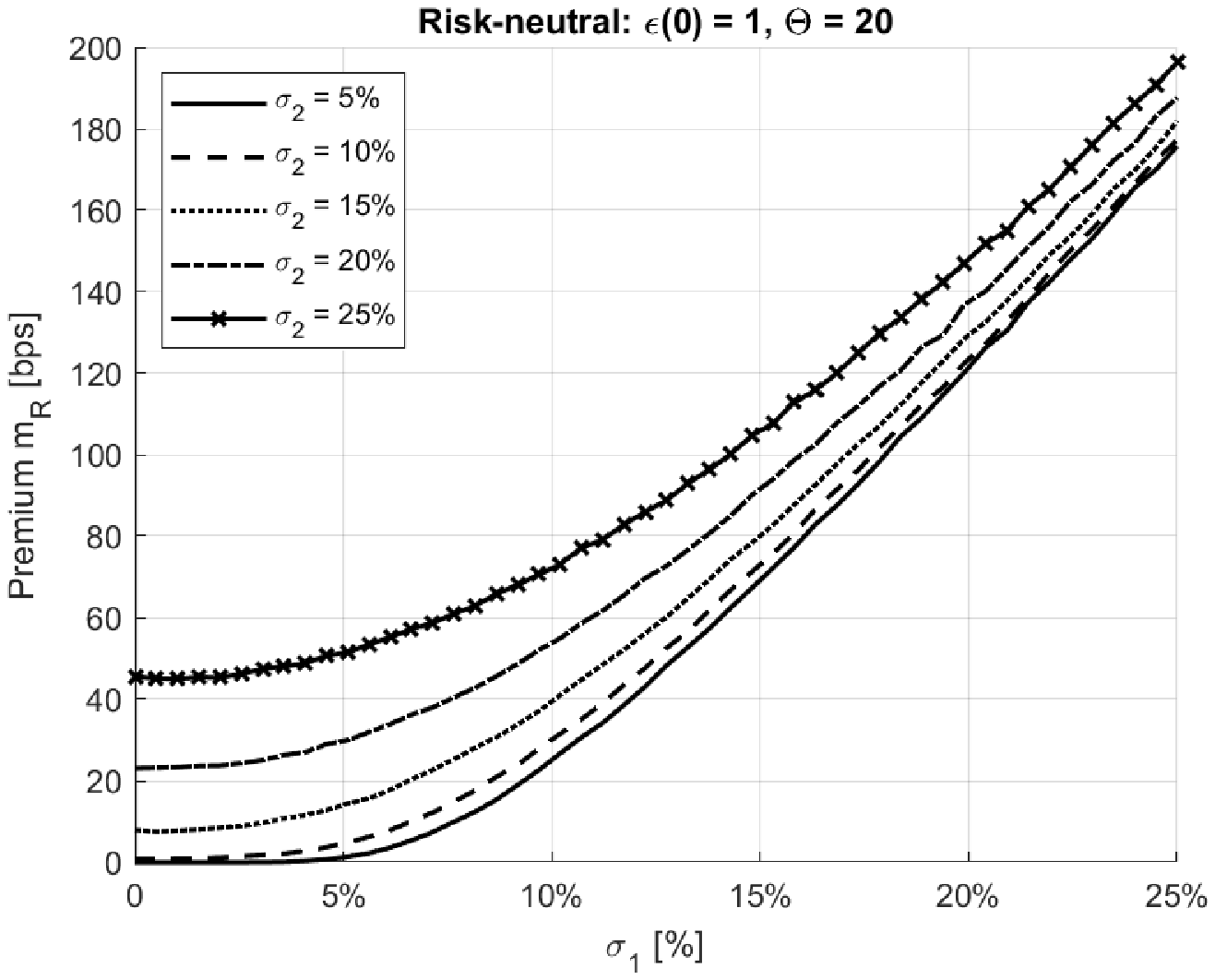}
\textit{This figure illustrates the value of the premium under risk-neutral valuation as a function of the volatility $\sigma_1$ for fixed levels of volatility $\sigma_2$ for different liquidation windows ($\Theta$ in days), when the state of the initial market is `good', i.e. $\tilde{\varepsilon}(0) = 1$. Parameter values are $T = 1$ (investment horizon), $X_0 = 1$ (initial investment), $c_M = 0.1$ (managerial deposit), $r = 0.01$ (risk-free rate), $p = 0.0175$ (probability of switching from `good' to 'stressed' state) and $q = 0.0865$ (probability of switching from `stressed' to `good' state).}
\caption{Numerical premium $m_R$ as a function of volatility - initial market `good'}
\label{MS (1)}
\end{figure}

\begin{figure}[H]
\includegraphics[width = 0.5\textwidth]{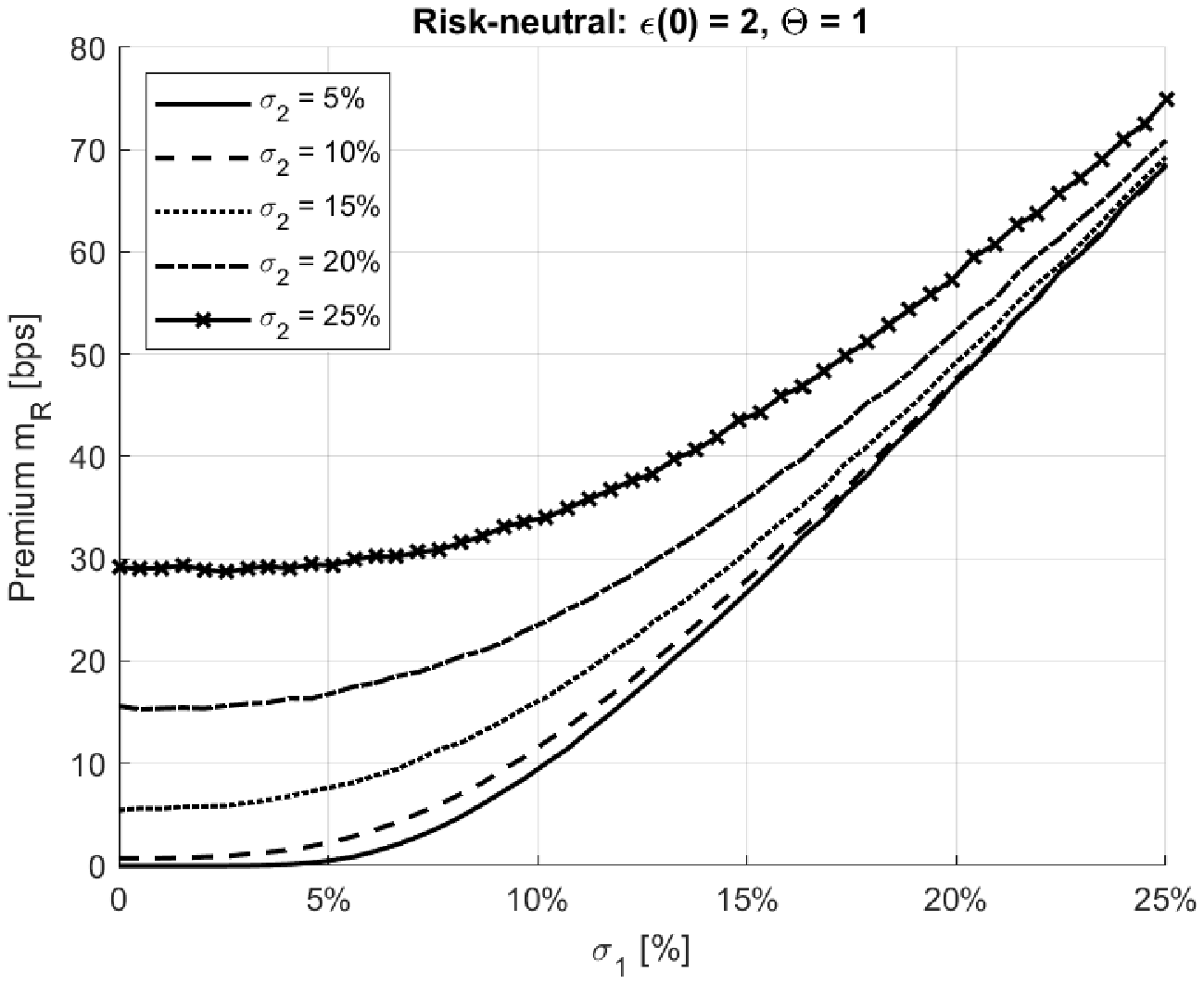}
\includegraphics[width = 0.5\textwidth]{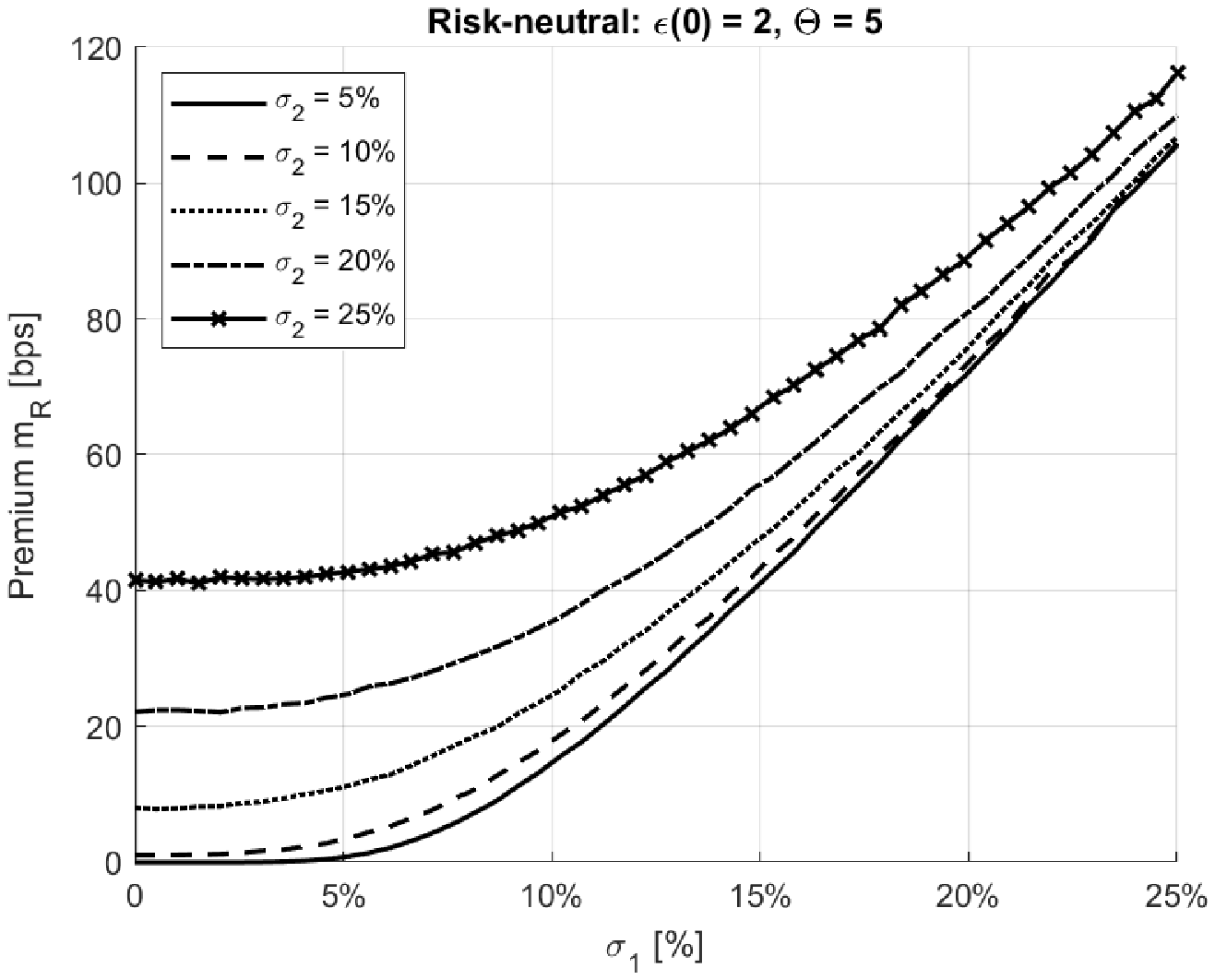} \\
\includegraphics[width = 0.5\textwidth]{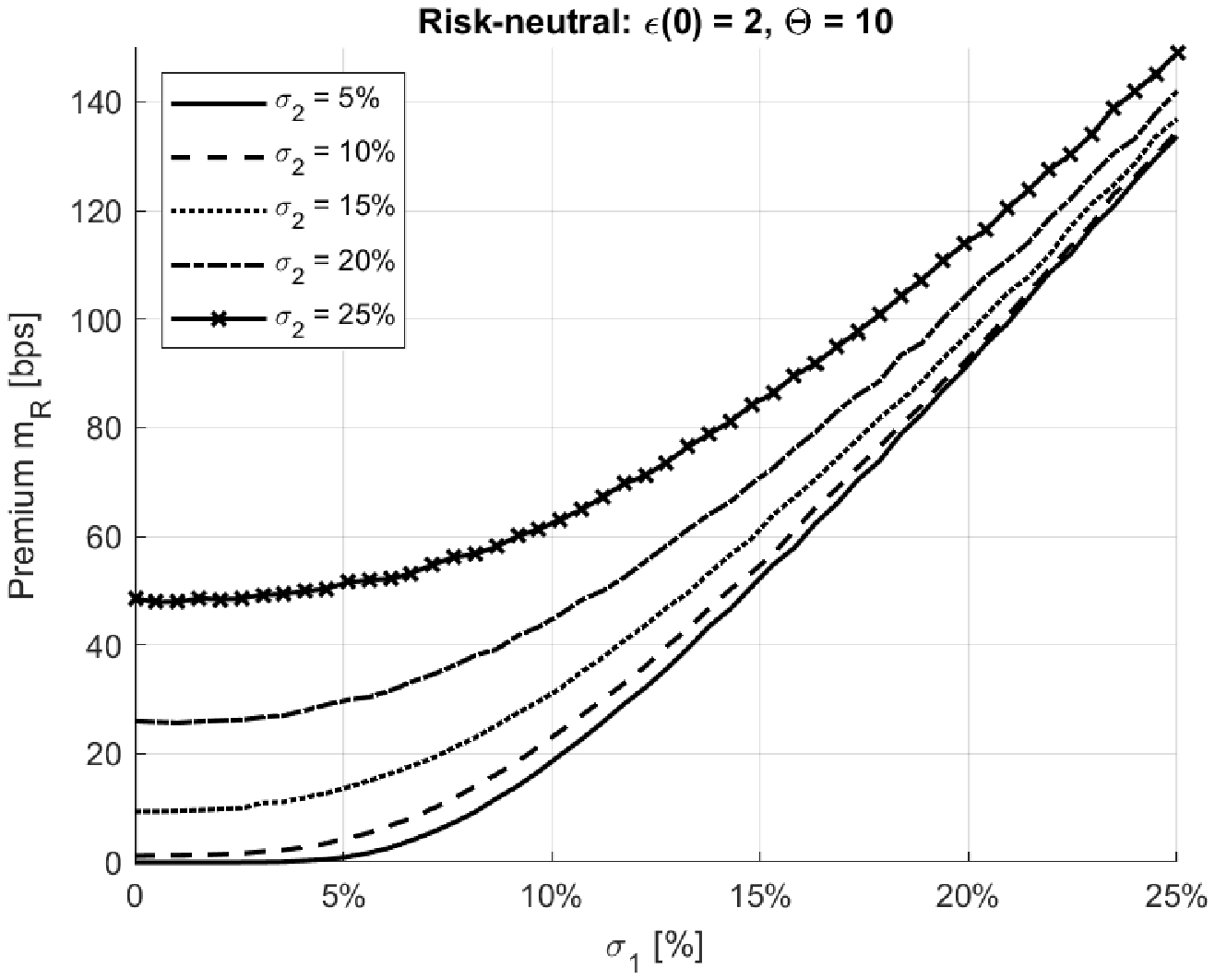}
\includegraphics[width = 0.5\textwidth]{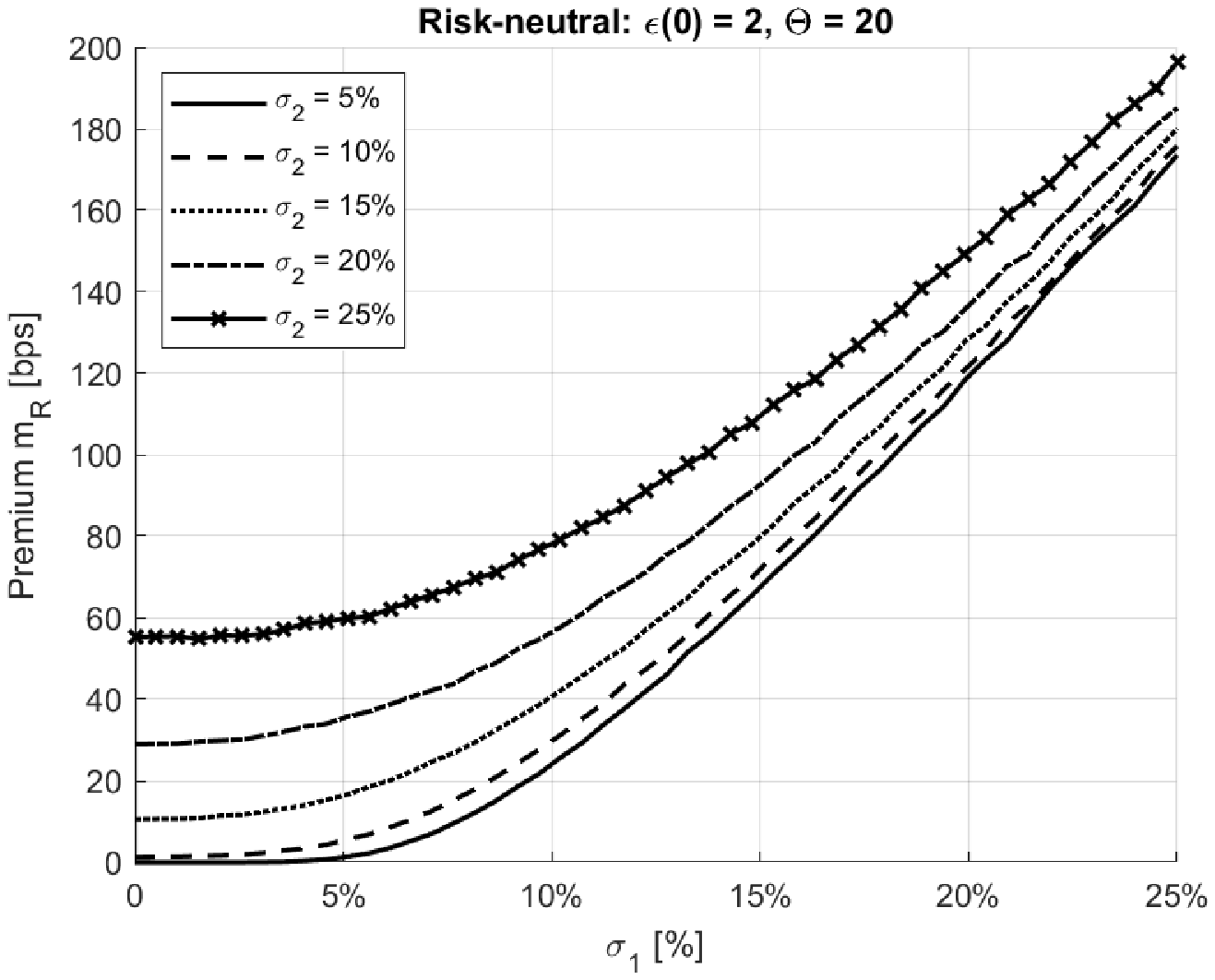}
\textit{This figure illustrates the value of the premium under risk-neutral valuation as a function of the volatility $\sigma_1$ for fixed levels of volatility $\sigma_2$ for different liquidation windows ($\Theta$ in days), when the state of the initial market is `stressed', i.e. $\tilde{\varepsilon}(0) = 2$. Parameter values are $T = 1$ (investment horizon), $X_0 = 1$ (initial investment), $c_M = 0.1$ (managerial deposit), $r = 0.01$ (risk-free rate), $p = 0.0175$ (probability of switching from `good' to `stressed' state) and $q = 0.0865$ (probability of switching from `stressed' to `good' state).}
\caption{Numerical premium $m_R$ as a function of volatility - initial market `stressed'}
\label{MS (2)}
\end{figure}

\begin{figure}[H]
{\centering \includegraphics[width = 0.75\textwidth]{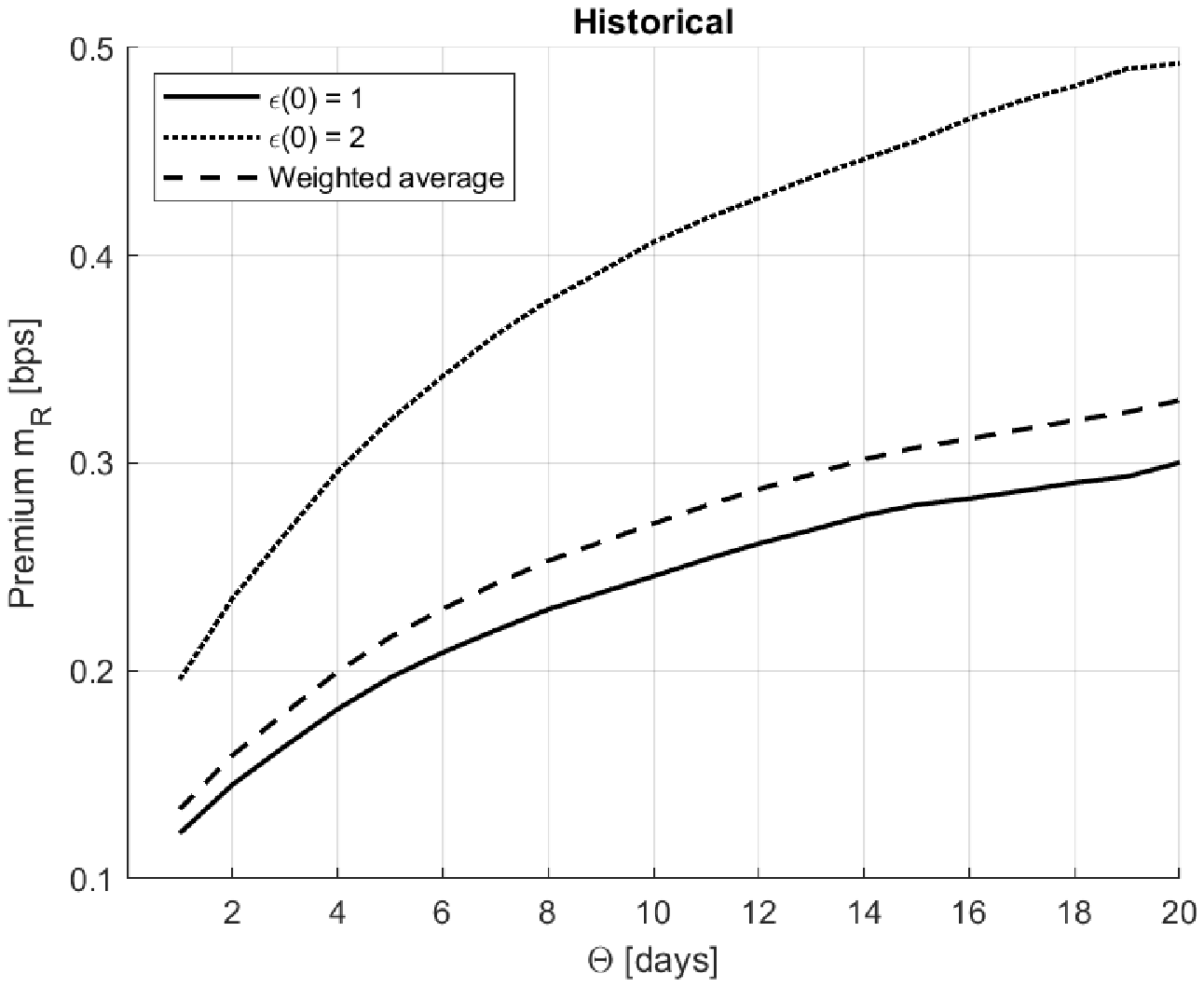}\\}
\textit{This figure illustrates the value of the premium under the discounted expected payoff valuation using empirical values as a function of the liquidity measure $\Theta$ (in days) for the empirical volatility $\sigma_1 = 0.0329$ and $\sigma_2 = 0.0895$ for both `good' initial market, i.e. $\tilde{\varepsilon}(0) = 1$, and `stressed' initial market, i.e. $\tilde{\varepsilon}(0) = 2$. The `weighted average' uses a probability of $0.8427$, that the market is in a `good' state. Parameter values are $T = 1$ (investment horizon), $X_0 = 1$ (initial investment), $c_M = 0.1$ (managerial deposit), $r = 0.01$ (risk-free rate), $\mu_1 = 0.0624$ (return `good' state), $\mu_2 = -0.1865$ (return `stressed' state), $p = 0.0175$ (probability of switching from `good' to `stressed' state) and $q = 0.0865$ (probability of switching from `stressed' to `good' state).}
\caption{Numerical premium $m_R$ as a function of volatility}
\label{MS (3)}
\end{figure}

Figures \ref{MS (1)} to \ref{MS (3)} show results for the numerical approach using a simulation with a Markov-switching model as underlying framework. The stationary distribution $\pi = (\pi_1,\pi_2) = (\frac{q}{p+q},\frac{p}{p+q}) = (0.8317,0.1683)$ for $p = 0.0175$ and $q = 0.0865$ indicates a high probability of the market being governed by state 1, i.e. the `good' market environment. Hence, Figures \ref{MS (1)} and \ref{MS (2)} illustrate the value of the premium depending on $\sigma_1$ for fixed levels of $\sigma_2$ and different levels of liquidity ($\Theta$ in days). Similar to results in the GBM section, figure \ref{MS (1)} (initial state is `good') illustrates the premium as a convex function of a fund's volatility. For low levels of volatility in both states ($\sigma_{1/2}\leq 5\%$) we find a marginal impact on the premium. However, assuming very liquid assets ($\Theta = 1$) even for extreme values in both states, i.e. $\sigma_{1/2} = 25\%$ the premium is significantly lower ($m_R \approx 74bps$) than in the single-regime model (for $\sigma = 25\%$ we find $m_R \approx 170bps$). For lower levels of volatility in state 1 ($\sigma_1 \leq 10\%$), the volatility of state 2 has much higher impact on the premium compared to a high volatility in state 1 ($\sigma_1 \approx 25\%$). Figure \ref{MS (2)} illustrates results for a market with initial state `stressed'. As expected, the premiums are higher than when assuming a `good' initial market. In figure \ref{MS (3)}, we see the premium using the discounted expected payoff valuation for empirical values depending on the initial state of the market. Assuming a `stressed' initial market, the premium is roughly twice as high as for a `good' initial market. Except for a daily liquidation window, even for `stressed' market, the premium is below the the premium using a single-regime model (see figure \ref{GBM (2)}). For a one-month liquidation window, premiums are approx. $0.3bps$ for a `good' market and approx. $0.5bps$ for a `stressed' market. For comparison: the premium for the same level of liquidity in the single-regime model is approx. $0.7bps$. The results indicate, that the selection of the applied model matters substantially for an investor. As the actual state of the market is not an observable variable, we also include a `weighted average'. The Baum-Welch algorithm indicates that the HFRX indices in table \ref{table HFRX} spent an average of $84.27\%$ in the `good' state during the period between April 1, 2003 and December 28, 2018. Using this number as a proxy for the probability, that the market is in a `good' state, we compute a `weighted average premium'.

\section{Backtesting}
\thispagestyle{empty}

\subsection{Setup and assumptions}
This Section elaborates on the backtesting of the suggested premium in the previous Sections. It aims to evaluate the practical application of the concept and how it would have performed in recent years. As the concept of insuring hedge fund portfolio losses beyond the first-loss tranche is completely new and no work has been done so far, there exist no predefined ways to evaluate the backtest or benchmark our results of the proposed premium to other approaches. Hence, this Section serves as a first starting point in improving the practical application of the concepts mentioned in this paper. The general idea of backtesting in this context is the evaluation of the computed premium using real world historical data. In order to obtain valid results, the look-ahead bias has to be avoided. This bias occurs when at certain times data or information, which was not available at the actual point in time of the event, is used for a simulation or evaluation of those events. The HFRX data set used for table \ref{table HFRX} is the main source of data used here. It contains daily returns from April 1, 2003 until December 28, 2018. Fund-depending parameters, e.g. returns, volatility and transition probabilities are determined using the HFRX data set. The annualized risk-free rates $r$ are extracted from the Kenneth R. French website\footnote{See http://mba.tuck.dartmouth.edu/pages/faculty/ken.french/data\_library.html for details}, which provides 1-month TBill returns starting in July, 1926. The following list describes the backtesting approach implement in this Section:
\begin{itemize}
\item[1)] Define a starting point, investment target (e.g. HFRXGL) and crucial parameters (i.e. liquidity measure $\Theta$, fees, managerial deposit $c_M$,initial investment $X_0$,...).
\item[2)] Estimate all parameters needed to compute the premium for a one-year insurance at the specific point in time depending on the predefined valuation approach, i.e. geometric Brownian motion framework or Markov-switching framework.
\item[3)] Determine the premium and thus the initial values for the investor, the manager and the reinsurer at the specific point in time.
\item[4)] Let the hedge fund performance develop according to the historical data.
\item[5a)] In the event that the fund value breaches the barrier at any time during the one-year investment horizon, all assets are liquidated and the payoff is evaluated using the final value depending on the underlying liquidity.
\item[5b)] In the event of no breach, evaluate the one-year performance at the end of the period.
\item[6)] (Only following step 5b)) Start a new investment period until the liquidation event occurs or a predefined stopping criterion is reached, e.g. when no historical data is available anymore.
\end{itemize}
Note, we are using the risk-neutral approach. The following elaborates how the specific parameters at a certain point in time are obtained and how the fund performance is evaluated. First, we illustrate the implemented algorithm: Let the initial investments for the investor $X_I(i)$, the manager $X_M(i)$ and the reinsurer $X_R(i)$ for investment period $i$, $i = 1,2,...,T-1$ be given by
\begin{align*}
X_I(i) &= \hat{V}_I(i-1) = X_i, \\
X_M(i) &= \hat{V}_M(i-1), \\
X_R(i) &= \hat{V}_R(i-1),
\end{align*}
where for $i=0$, the following values are set: $X_I(0) = X_0$, $X_M(0) = 0$ and $X_R(0) = 0$. Note, in this paper $X_0 = 1$. At the beginning of the investment period $i$ the premium $m_{R_i}$ is computed and the initial values for the investor $V_I(i)$, the manager $V_M(i)$ and the reinsurer $V_R(i)$ for investment period $i$, $i = 0,1,...,T-1$ are given by
\begin{align*}
V_I(i) &= (1-m_{R_i}) X_I(i),\\
V_M(i) &= X_M(i), \\
V_R(i) &= X_R(i) + m_{R_i} X_I(i).
\end{align*}
The final values for the investor $\hat{V}_I(i)$, the manager $\hat{V}_M(i)$ and the reinsurer $\hat{V}_R(i)$ for investment period $i$, $i = 0,1,...,T-1$ are given by
\begin{align*}
\hat{V}_I(i) &= X_{i,i+1}-m_{R_i} e^r X_I(i) - \alpha_M[X_{i,i+1}-X_I(i)]^+ + [X_I(i) - X_{i,i+1}]^+, \\
\hat{V}_M(i) &= e^r V_M(i) + \alpha_M[X_{i,i+1}-X_I(i)]^+ - [X_I(i) - X_{i,i+1}]^+ +[(1-c_M)X_I(i)-X_{i,i+1}]^+, \\
\hat{V}_R(i) &= e^r V_R(i) - [(1-c_M)X_I(i)-X_{i,i+1}]^+,
\end{align*}
where $X_{i,i+1}$ denotes the one-year price development of the initial investment $X_I(i)$ in period $i$, hence it is the final value of the initial investment $X_I(i)$ for this period, which is used as initial investment in period ($i+1$). Note, for a one-period investment horizon and no flat management fee ($m_M = 0$), the equations above yield equations (\ref{payoff I new}) to (\ref{payoff R new}). Now, if for any $\tau \in [0,1]$ it holds that $X_{i,i+\tau}\leq (1-c_M)X_i=(1-c_m)X_I(i)$ the hedge fund is liquidated depending on $\Theta$, e.g. in this paper $\Theta \in \{\frac{1}{252}, \frac{5}{252}, \frac{10}{252}, \frac{20}{252}\}$, and the final value is set to the following:
\begin{align*}
\hat{V}_I(i) = \, &X_{i,i+\tau+\Theta}-m_{R_i} e^{r(\tau+\Theta)} X_I(i) - \alpha_M[X_{i,i+\tau+\Theta}-X_I(i)]^+ + [X_I(i) - X_{i,i+\tau+\Theta}]^+, \\
\hat{V}_M(i) = \, &e^{r(\tau+\Theta)} V_M(i) + \alpha_M[X_{i,i+\tau+\Theta}-X_I(i)]^+ - [X_I(i) - X_{i,i+\tau+\Theta}]^+\\
 &+[(1-c_M)X_I(i)-X_{i,i+\tau+\Theta}]^+, \\
\hat{V}_R(i) = \,&e^{r(\tau+\Theta)} V_R(i) - [(1-c_M)X_I(i)-X_{i,i+\tau+\Theta}]^+.
\end{align*}
Note, we employ daily time steps in our backtest. This allows for a smoother performance graph.

As already mentioned, the approach is a simplification and several assumptions and specifications are made:
\begin{itemize}
\item[-] Rolling windows: Premiums are computed based on parameters obtained from a two-year rolling window. In the geometric Brownian motion approach, the estimated volatility is simply the sample volatility. For the Markov-switching approach, we use the Baum-Welch algorithm according to Section \ref{section Valuation in a Markov-switching framework} to estimate volatility and transition probabilities.
\item[-] No reinvestment: This applies to the manager and reinsurer only. All premiums and performance fees are not reinvested in the hedge fund but transferred to a riskless investment account. In other words, the upfront premiums are compounded using the risk-free rate at the time of occurrence. Premiums from following investment periods are added to the account and compounded until either the liquidation event is triggered or the ultimate investment horizon is reached. We are aware of the fact that in practice the premiums collected by an insurance company are subject to individual asset/liability management (ALM) processes. In some cases it is even plausible to reinvest parts or the entire premium in the fund and participate in upside performance. As this would overly complicate inter-party relationships, we implement the simpler approach. The value of the manager follows similar assumptions. In practice it is common that the manager owns a stake of the fund and reinvests performance fees. In our approach, all performance fees are paid at the end of the investment period and are transferred to a riskless investment account and compounded.
\item[-] No consumption: The investor is reinvesting all available funds in the following investment period, i.e. the final value of investment period $t$ is the initial investment of period $t+1$. During the investment period, the investor cannot redeem her stake. In practice, this is enforced by lock-up periods, however, they usually only last up to two years. The manager's value is not subject to any consumption either. In practice, the manager is taking out funds according to either a predefined schedule or at will (e.g. to pay staff salaries, etc.). In our approach, the manager does not own a personal stake in the initial fund and compounded performance fees are even used to offset potential losses in the liquidation event. The same argument holds for the reinsurer and is linked to the no reinvestment assumption above.
\item[-] Investment horizon: We implement consecutive independent one-year investment periods. All fees are paid annually: the premium is paid upfront and the performance fee is paid at the end of the respective investment period. The performance fee is obtained according to a predefined schedule and in the course of the one-year period. The investment periods are independent from each other, i.e. every year is implemented as if it was a standalone investment and results are accumulated.
\item[-] Hurdle rates: The liquidation barrier is reset at the beginning of each period and is only valid for this period. With the beginning of the following period, the liquidation barrier is reset depending on the new initial investment. In practice, many hedge funds have hurdle rates or high watermarks in place, which we ignore in this simplified approach.
\item[-] Buy and hold strategy: In case a portfolio is made up of multiple indices, we implement a simple buy and hold strategy. A fictitious initial portfolio with equally weighted indices is the basis for the calculation of the portfolio's (weighted) returns and volatility, i.e. during the investment periods, weights are adjusted depending on the individual performance of the underlying indices when calculating returns and not kept constant. There is no dynamic rebalancing at any time.
\item[-] Time schedule: The starting date for any portfolio is April 1, 2005. This allows for two complete years of daily return data for the first rolling window. Each year on April 1 a new premium is computed and the last trading day (and the evaluation of the portfolio) for this period is March 31. If no liquidation process is triggered, April 1, 2017 is the last start date for a one-year period. In total this allows a maximum overall investment horizon of 13 years (April 1, 2005 until March 31, 2018).
\end{itemize}

\subsection{Results}
\begin{figure}[H]
\includegraphics[width = \textwidth]{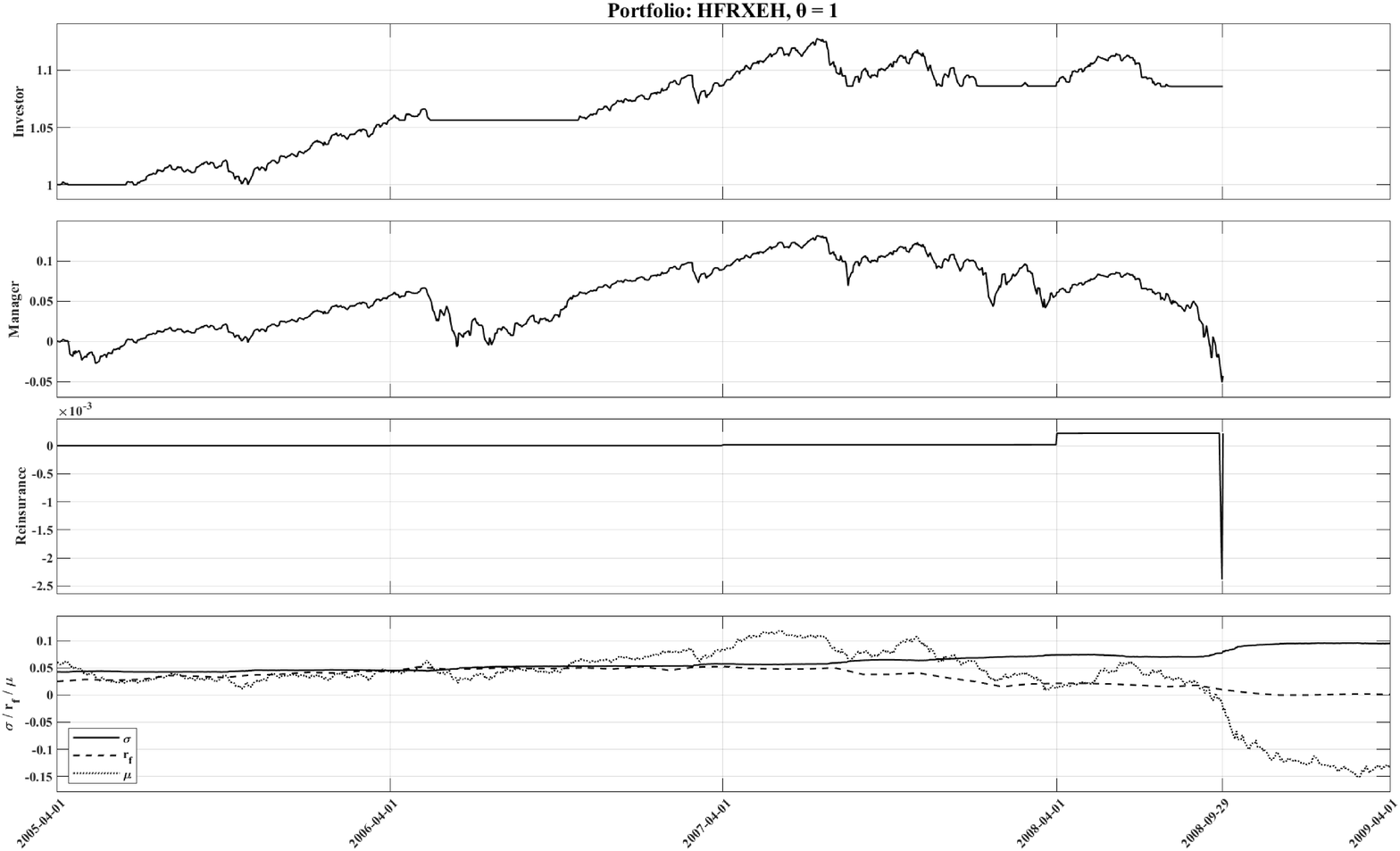}
\textit{This figure illustrates the HFRXEH backtesting performance for $\Theta = 1$ (in days) using the GBM approach. Barrier breach: September 29, 2008.}
\caption{GBM HFRXEH backtesting (1)} 
\label{backtest1}
\includegraphics[width = \textwidth]{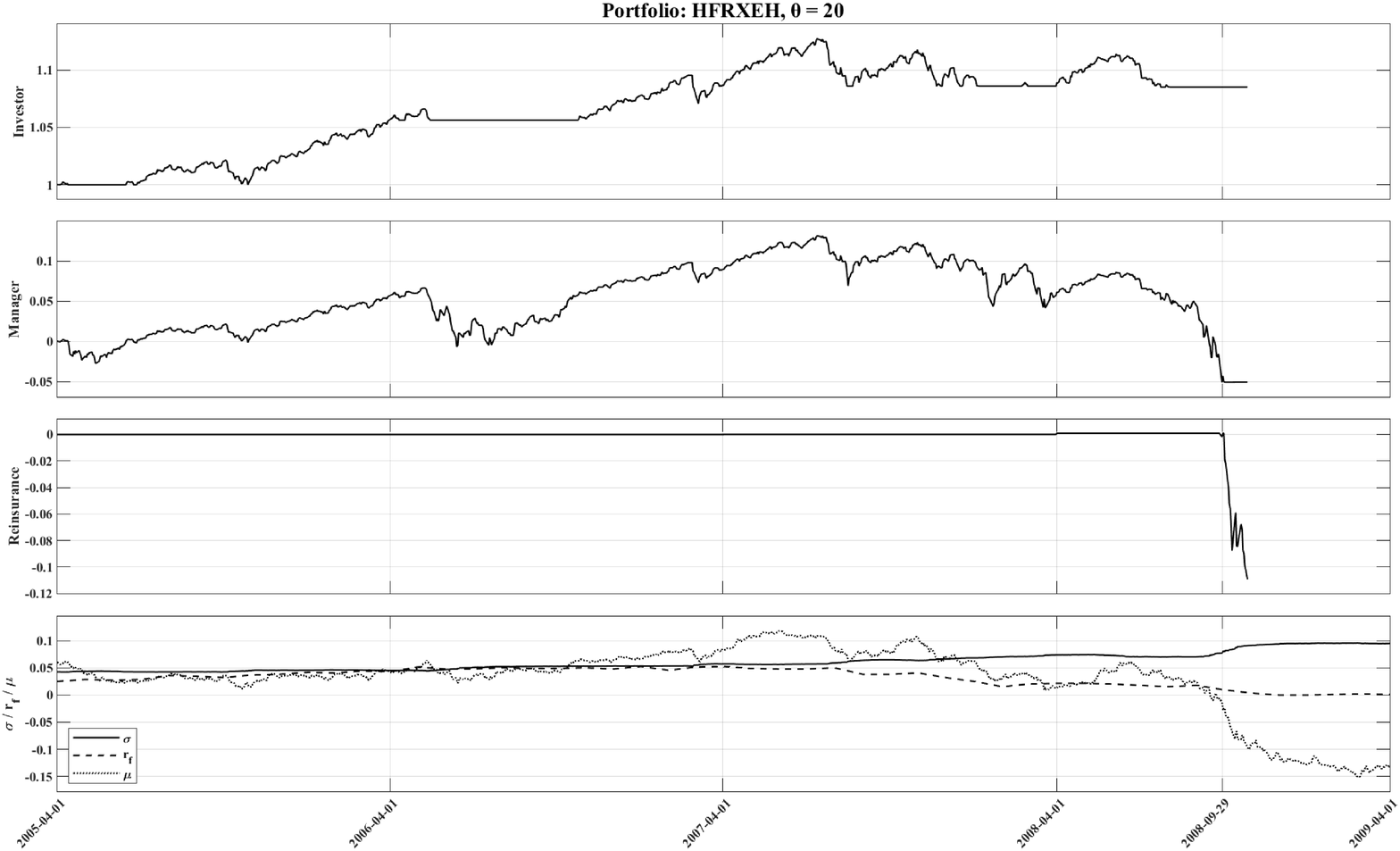}\\
\textit{This figure illustrates the HFRXEH backtesting performance for $\Theta = 20$ (in days) using the GBM approach. Barrier breach: September 29, 2008.}
\caption{GBM HFRXEH backtesting (2)}
\label{backtest2}
\end{figure}

\begin{figure}[H]
\includegraphics[width = \textwidth]{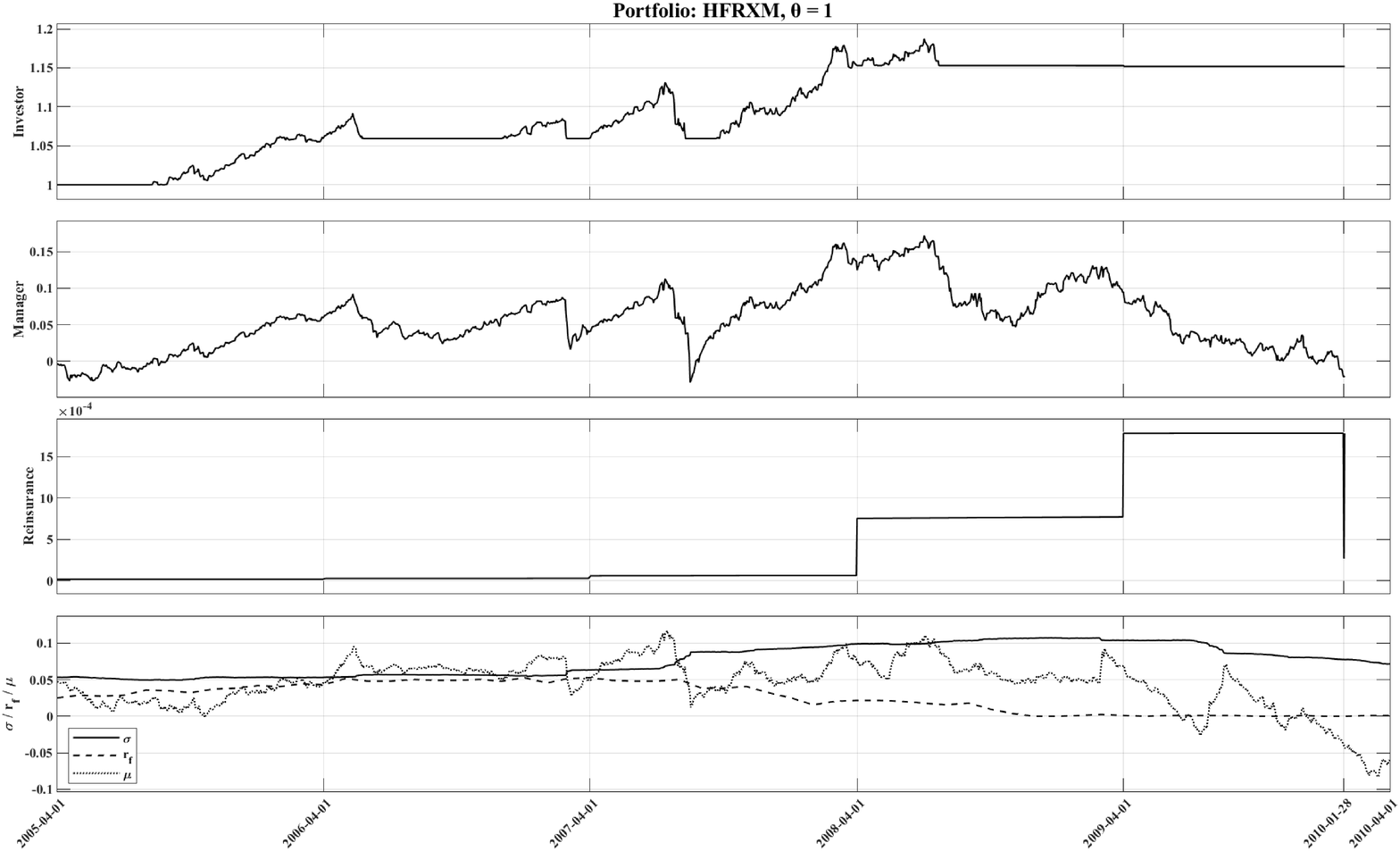}
\textit{This figure illustrates the HFRXM backtesting performance for $\Theta = 1$ (in days) using the GBM approach. Barrier breach: January 28, 2010.}
\caption{GBM HFRXM backtesting (1)} 
\label{backtest3}
\includegraphics[width = \textwidth]{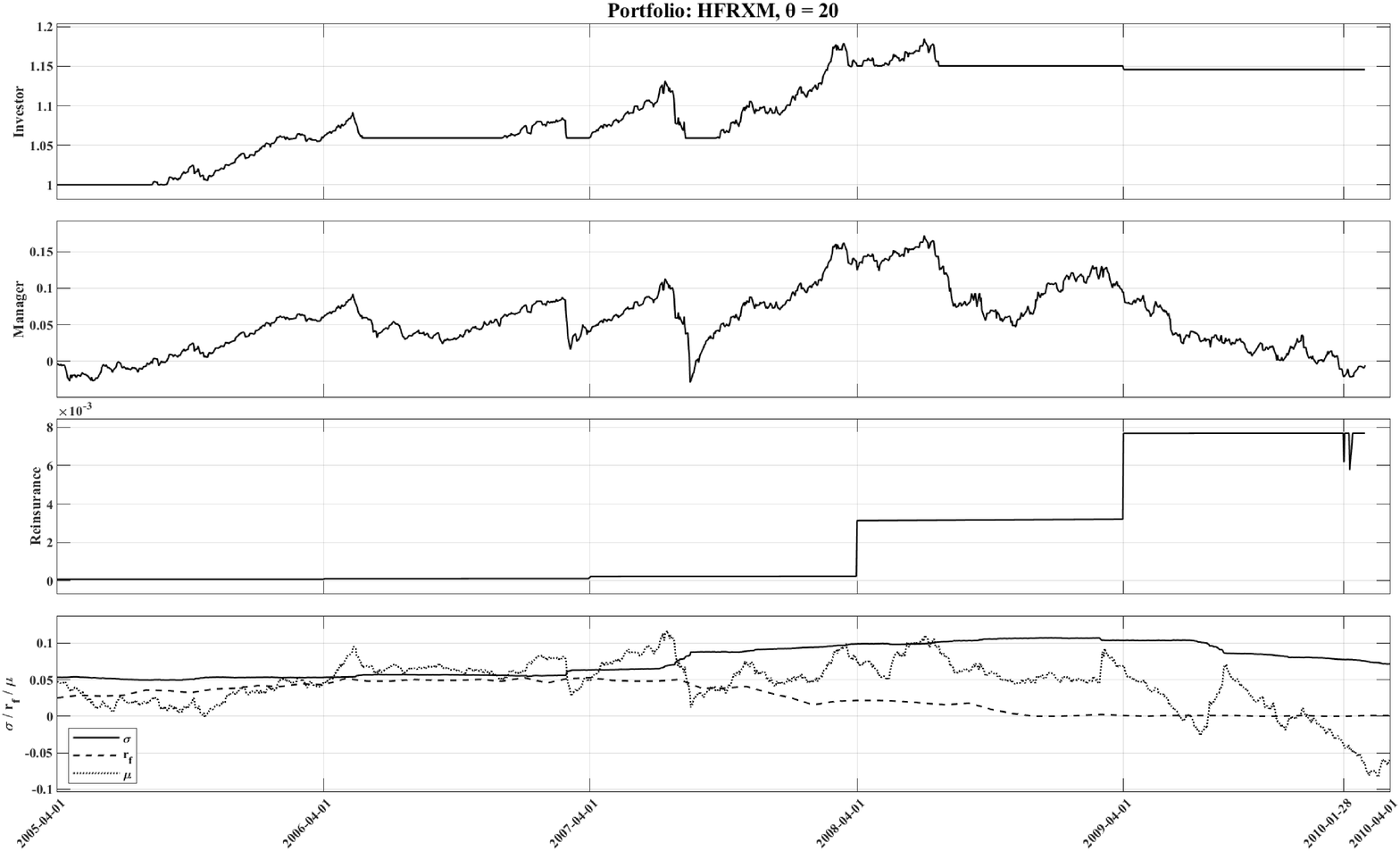}\\
\textit{This figure illustrates the HFRXM backtesting performance for $\Theta = 20$ (in days) using the GBM approach. Barrier breach: January 28, 2010.}
\caption{GBM HFRXM backtesting (2)}
\label{backtest4}
\end{figure}

\begin{figure}[H]
\includegraphics[width = \textwidth]{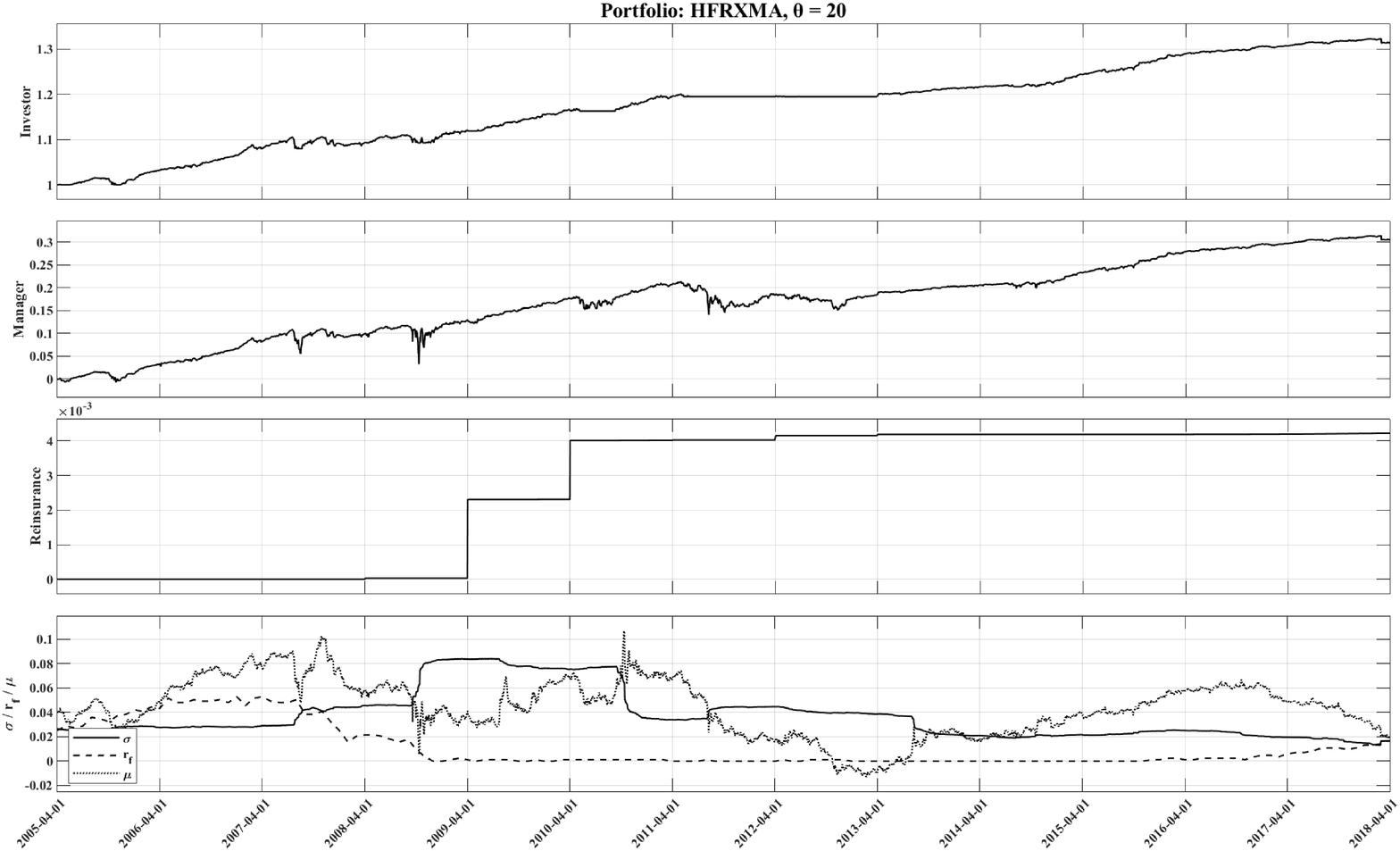}
\textit{This figure illustrates the HFRXGL backtesting performance for $\Theta = 20$ (in days) using the GBM approach. No barrier breach occurred.}
\caption{GBM HFRXGL backtesting} 
\label{backtest5}
\includegraphics[width = \textwidth]{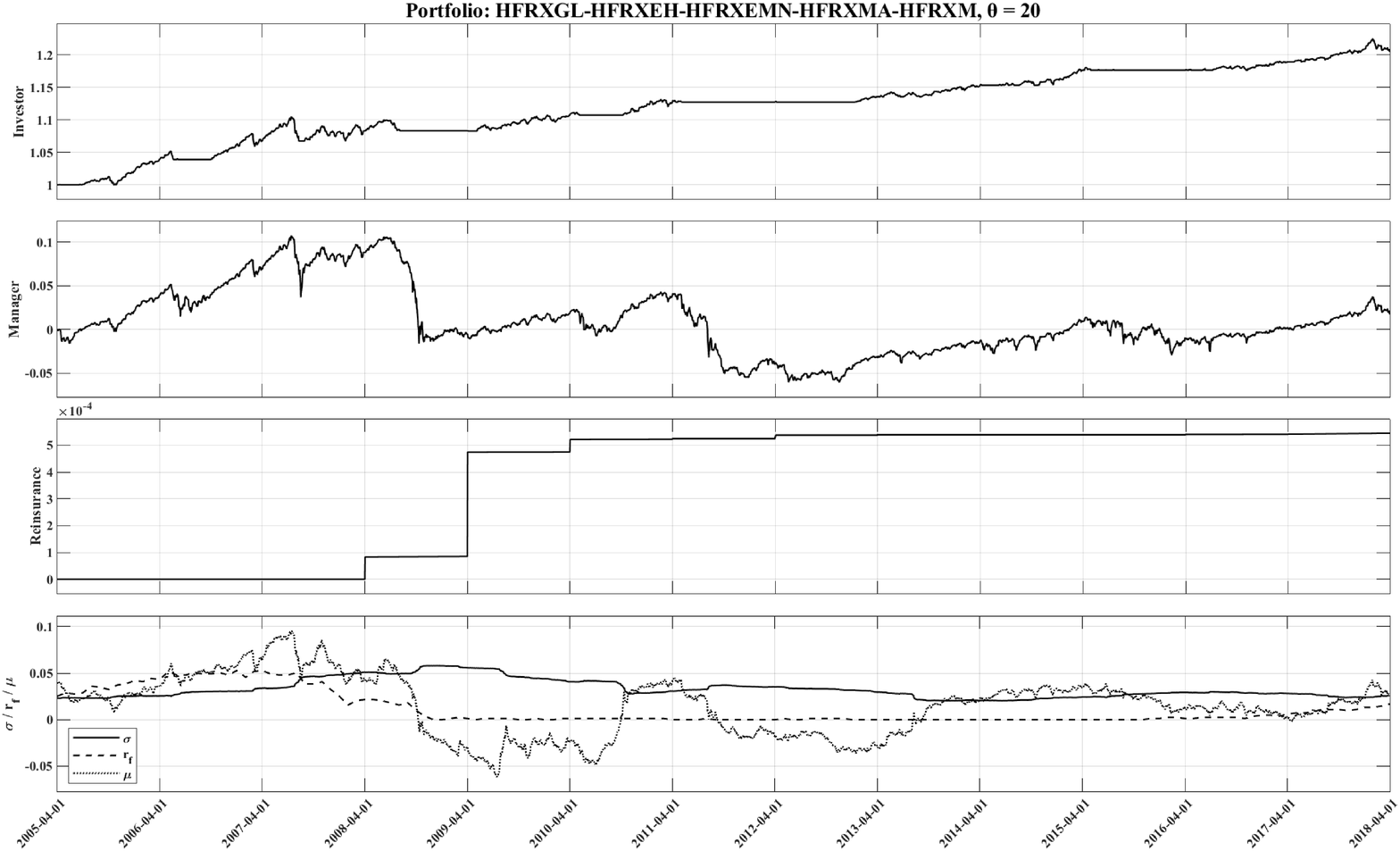}\\
\textit{This figure illustrates the backtesting performance for an equally weighted (all HFRX indices) portfolio for $\Theta = 20$ (in days) using the GBM approach. No barrier breach occurred.}
\caption{GBM HFRX portfolio backtesting}
\label{backtest6}
\end{figure}

Note that in the following we refer to the units of $\Theta$ in days, i.e. $\Theta = 1$ represents a one-day liquidation window, which is implemented in the simulation using $\Theta = \frac{1}{252}$. Figures \ref{backtest1} to \ref{backtest6} show results of the implemented backtesting using the geometric Brownian motion valuation according to Section \ref{Analytic Valuation using Geometric Brownian Motion}. Figures \ref{backtest1} and \ref{backtest2} illustrate the HFRX Equity Hedge Index (HFRXEH) for $\Theta = 1$ and $\Theta = 20$. Whereas in figure \ref{backtest1} the reinsurer shows a profit for $\Theta = 1$, the reinsurer shows a massive deficit for $\Theta = 20$. After breaching the liquidation barrier, the price of the underlying fund shortly increased but heavily decreased over the following period. In this figure, two features are graphically visible: First, during any investment period (illustrated by the area between two ticks on the time-axis), the investor's value never drops below the respective period's initial value. In other words, the initial value is the lower limit for the investor's value in any period. This is illustrated by the flat parts in the top graph. Second, note the cap on the manager's obligations, illustrated by the flat line in the second graph at the end. Here, the reinsurance is triggered. Figures \ref{backtest3} and \ref{backtest4} illustrate the HFRX Macro\textbackslash CTA Index (HFRXM) and represent examples, where the reinsurer is able to cover losses beyond the first-loss tranche using the collected premiums. In figure \ref{backtest3} ($\Theta = 1$), all losses are offset by the premium. In figure \ref{backtest4} ($\Theta = 20$) the fund value actually rises above the level of the managerial deposit, i.e. the reinsurance is not held liable, however, the collected premiums would have offset occurring losses at any time between the barrier breach and full liquidation. Also, note the difference in the multiplier due to the increase in $\Theta$: for $\Theta = 1$ the value of the reinsurer reaches a maximum of $V_R \approx 18 \cdot 10^{-4} = 18bps$ and for $\Theta = 20$ a maximum of $V_R \approx 7.5 \cdot 10^{-3} = 75bps$. Figure \ref{backtest5} shows the HFRX ED: Merger Arbitrage Index (HFRXMA), where during the entire investment horizon the liquidation process is not triggered. Here, the investor steadily builds up her value. Note the jump in the premium for the reinsurance, when markets are very volatile (illustrated by rising volatility in the bottom graph) in April 2009 and 2010. In figure \ref{backtest6} we implemented an equally weighted portfolio of all five HFRX indices listed in table \ref{table HFRX}. Due to the diversification, there is no liquidation breach even though some indices show a massive negative performance during the financial crisis.

\begin{figure}[H]
\includegraphics[width = \textwidth]{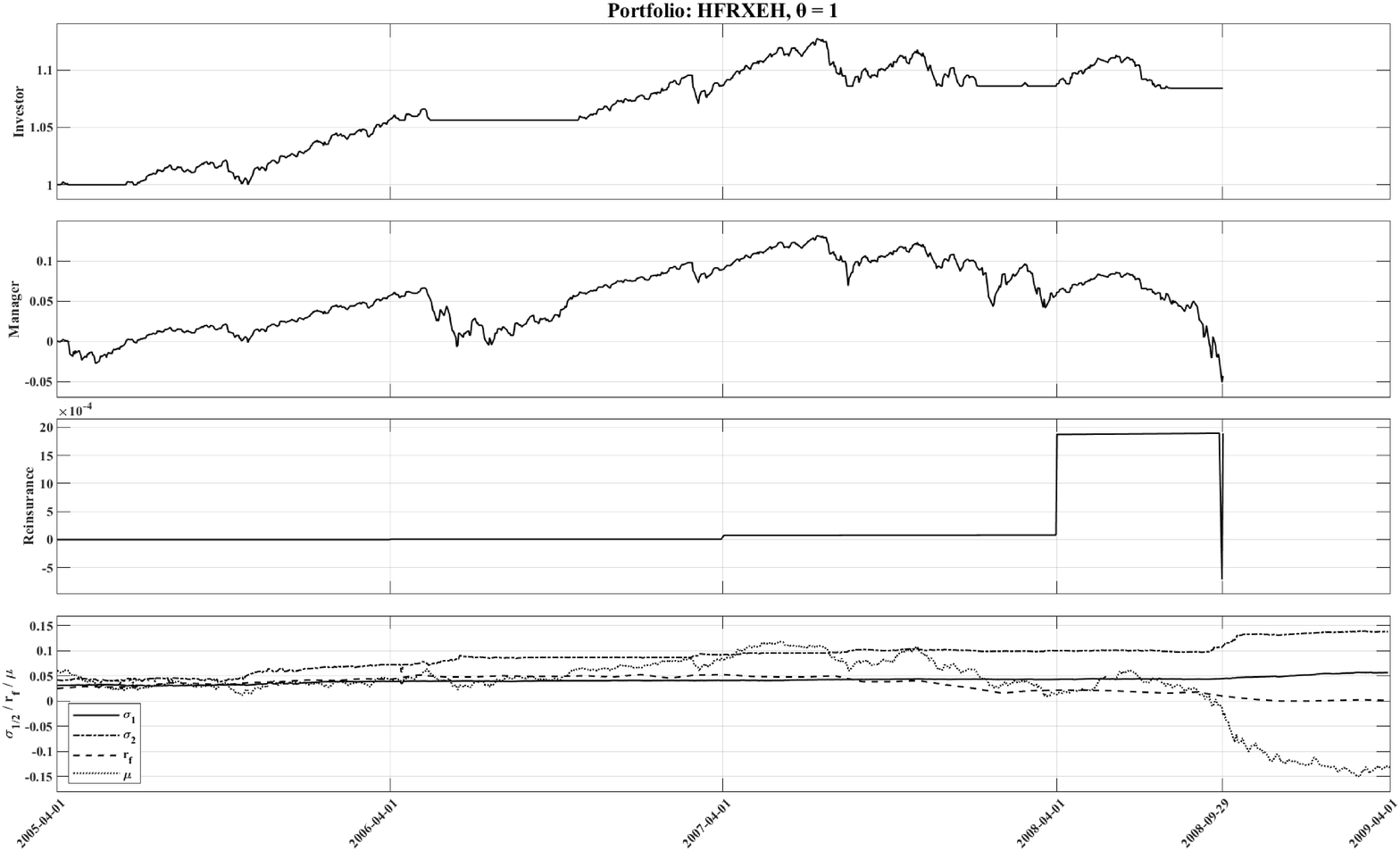}
\textit{This figure illustrates the HFRXEH backtesting performance for $\Theta = 1$ (in days) using the MS approach. Barrier breach: September 29, 2008.}
\caption{MS HFRXEH backtesting (1)} 
\label{backtest7}
\includegraphics[width = \textwidth]{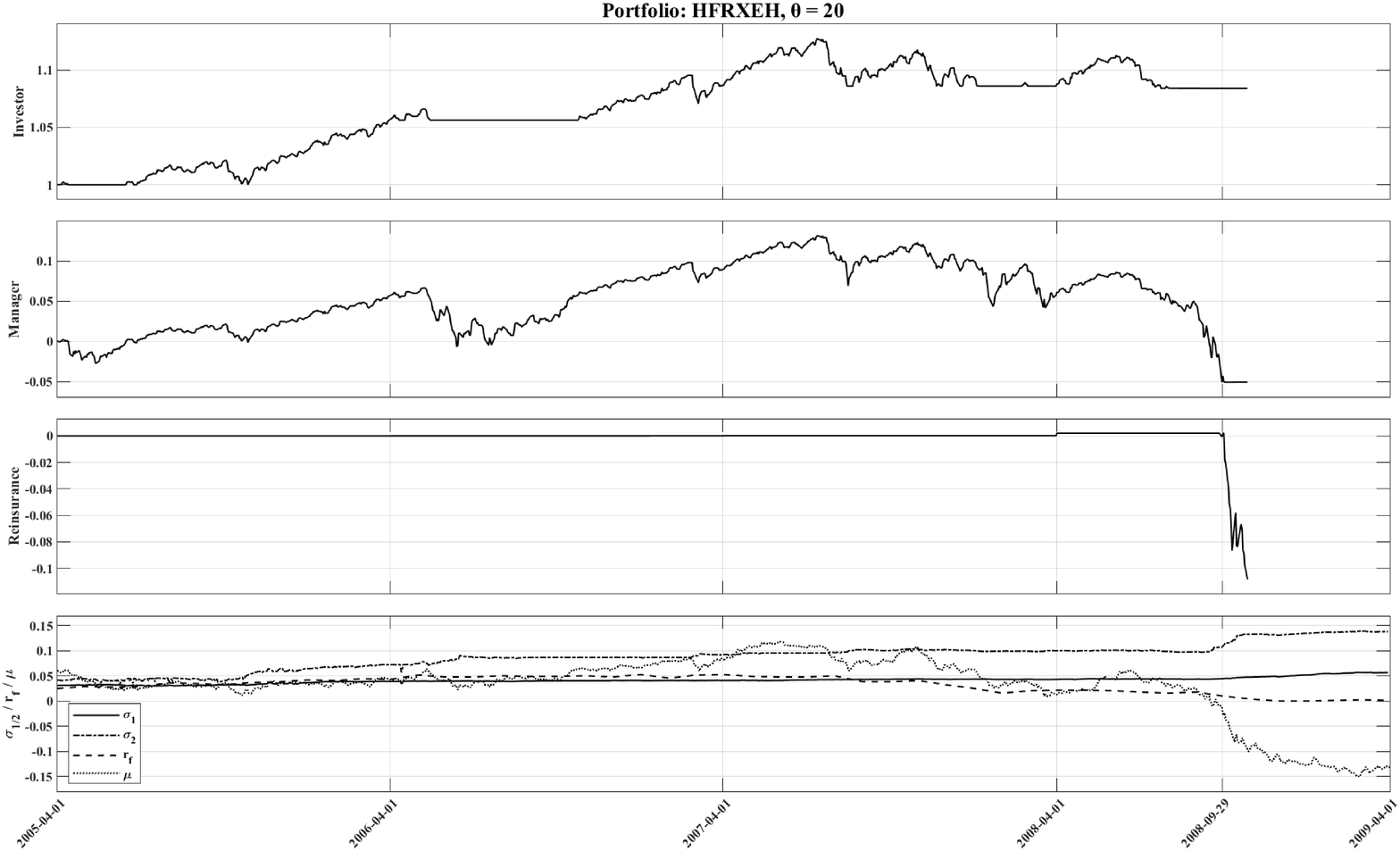}\\
\textit{This figure illustrates the HFRXEH backtesting performance for $\Theta = 20$ (in days) using the MS approach. Barrier breach: September 29, 2008.}
\caption{MS HFRXEH backtesting (2)}
\label{backtest8}
\end{figure}

\begin{figure}[H]
\includegraphics[width = \textwidth]{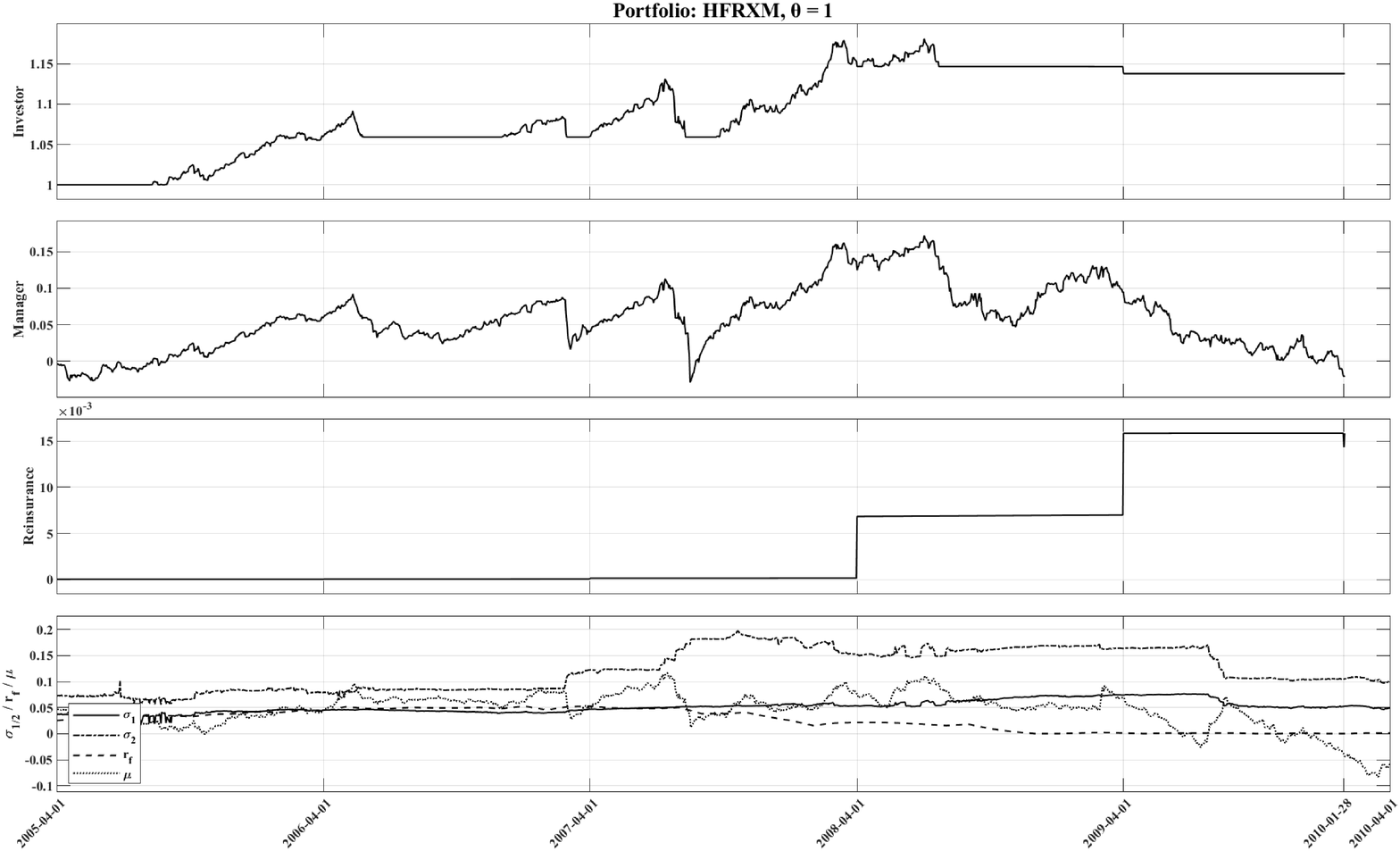}
\textit{This figure illustrates the HFRXM backtesting performance for $\Theta = 1$ (in days) using the MS approach. Barrier breach: January 28, 2010.}
\caption{MS HFRXM backtesting (1)} 
\label{backtest9}
\includegraphics[width = \textwidth]{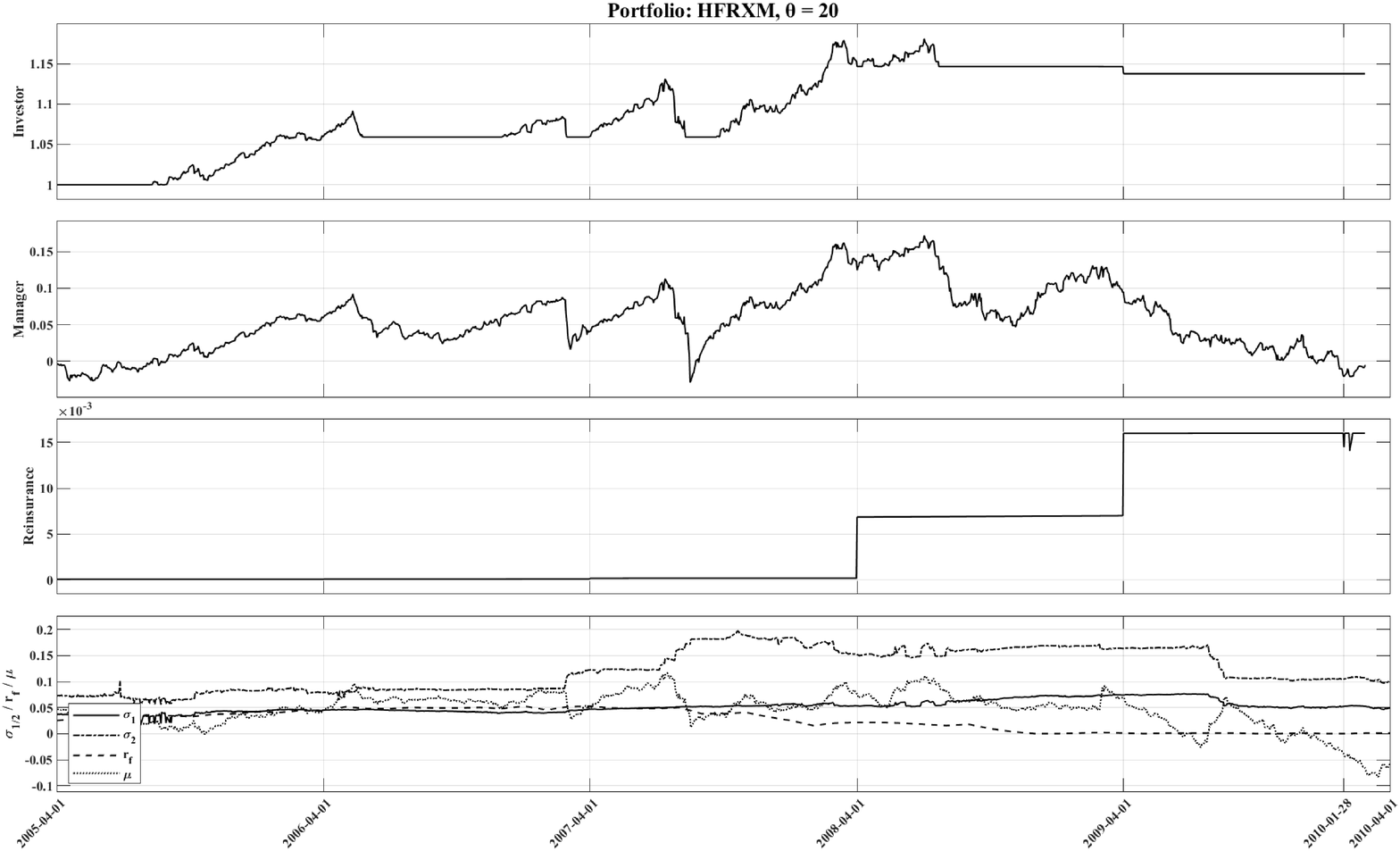}\\
\textit{This figure illustrates the HFRXM backtesting performance for $\Theta = 20$ (in days) using the MS approach. Barrier breach: January 28, 2010.}
\caption{MS HFRXM backtesting (2)} 
\label{backtest10}
\end{figure}

Figures \ref{backtest7} to \ref{backtest10} show results for the Markov-switching approach according to Chapter \ref{section Valuation in a Markov-switching framework}. Figures \ref{backtest7} and \ref{backtest8} show roughly the same results for the HFRXEH Index as figures \ref{backtest1} and \ref{backtest2} (GBM approach) except that the amount of the premium is higher for the Markov-switching approach. Figures \ref{backtest10} (MS) and \ref{backtest4} (GBM) yield similar results for the HFRXM Index. However, the maximum value of the reinsurer in the Markov-switching model is more than twice as high ($V_R \approx 16 \cdot 10^{-3} = 160bps$), due to higher premiums.

Summing up the above: We obtain mixed results for both the geometric Brownian motion and the Markov-switching approach. In some cases, losses can be absorbed by the paid premium. However, especially during the financial crisis, when sudden crashes caused funds to suffer heavy losses, the premium was by far not high enough.


\section{Conclusion}
We suggest an upfront premium to a reinsurance party in exchange for full portfolio protection, which goes beyond the insurance of the first tranche of losses by a first-loss scheme. The additional downside protection extends the first tranche by considering a second one, that will insure the investor against all losses, not just the first level. In this paper we propose both an analytical closed-form solution based on a derivative-pricing approach with a geometric Brownian motion as underlying framework and a numerical solution using a simulation based on a Markov-switching model.  In both approaches, we assume liquidity to be the key parameter and thus, the premium is derived as a function of the underlying fund's liquidity. A simplified backtesting method delivers mixed results: for some hedge fund indices, historical losses especially during the financial crisis were too high to be covered by the calculated premium. However, in some cases the reinsurer was able to offset losses by collected premiums. 

There are several extensions and questions that could be subject to future research. In this paper, we suggest a total of two tranches. This framework could be extended, creating a fund structure similar to an asset-backed security containing several tranches, each with a different risk and return profile. The parameters could also be altered to a point where both the reinsurance and the manager earn a fixed fee and a performance dependent fee. Finally, the fund's underlying liquidity is assumed to be homogeneous with daily liquidation steps in our approach. Combining our approach with certain liquidation time distributions, e.g. an exponential or Weibull distribution as suggested in \textcite{bordag2014portfolio} and the optimization of the liquidation process that comes along with the issue could be considered. 


\addcontentsline{toc}{chapter}{References}
\printbibliography

\end{document}